\documentclass[onecolumn, numberedappendix]{aastex6}
\usepackage{color}

\newcommand{\eg}{{\it e.g.}}
\newcommand{\etal}{et~al.}

\newcommand{\ks}{$K_{\rm s}$}
\newcommand{\vmk}{$(V-K_{\rm s})_0$}

\newcommand{\msun}{M$_{\sun}$}

\newcommand{\teff}{$T_{\rm eff}$}
\newcommand{\vsini}{$v \sin i$}

\begin{document}

\title{Rotation in the Pleiades with K2: I. Data and First Results }

\slugcomment{Version from \today}

\author{L.~M.~Rebull\altaffilmark{1,2},
J.~R.~Stauffer\altaffilmark{2},
J.~Bouvier\altaffilmark{3}, 
A.~M.~Cody\altaffilmark{4},
L.~A.~Hillenbrand\altaffilmark{5}, 
D.~R.~Soderblom\altaffilmark{6}, 
J.~Valenti\altaffilmark{6}, 
D.~Barrado\altaffilmark{7},
H.~Bouy\altaffilmark{7},
D.~Ciardi\altaffilmark{8},
M.~Pinsonneault\altaffilmark{9},
K.~Stassun\altaffilmark{10},
G.~Micela\altaffilmark{11},
S.~Aigrain\altaffilmark{12},
F.~Vrba\altaffilmark{13},
G.~Somers\altaffilmark{9},
J.~Christiansen\altaffilmark{8},
E.~Gillen\altaffilmark{12,14},
A.~Collier Cameron\altaffilmark{15}}

\altaffiltext{1}{Infrared Science Archive (IRSA), Infrared Processing
and Analysis Center (IPAC), 1200 E.\ California Blvd., California
Institute of Technology, Pasadena, CA 91125, USA; rebull@ipac.caltech.edu}
\altaffiltext{2}{Spitzer Science Center (SSC), Infrared Processing and
Analysis Center (IPAC), 1200 E.\ California Blvd., California
Institute of Technology, Pasadena, CA 9112, USA5}
\altaffiltext{3}{Universit\'e de Grenoble, Institut de Plan\'etologie
et d'Astrophysique de Grenoble (IPAG), F-38000 Grenoble, France;
CNRS, IPAG, F-38000 Grenoble, France}
\altaffiltext{4}{NASA Ames Research Center, Kepler Science Office,
Mountain View, CA 94035, USA}
\altaffiltext{5}{Astronomy Department, California Institute of
Technology, Pasadena, CA 91125, USA}
\altaffiltext{6}{Space Telescope Science Institute, 3700 San Martin Drive,
Baltimore, MD 21218, USA; Center for Astrophysical Sciences, Johns Hopkins University,
3400 North Charles St., Baltimore, MD 21218, USA}
\altaffiltext{7}{Centro de Astrobiolog\'ia, Dpto. de
Astrof\'isica, INTA-CSIC, E-28692, ESAC Campus, Villanueva de
la Ca\~nada, Madrid, Spain}
\altaffiltext{8}{NASA Exoplanet Science Institute (NExScI), Infrared
Processing and Analysis Center (IPAC), 1200 E.\ California Blvd.,
California Institute of Technology, Pasadena, CA 91125, USA}
\altaffiltext{9}{Department of Astronomy, The Ohio State University,
Columbus, OH 43210, USA; Center for Cosmology and Astroparticle
Physics, The Ohio State University, Columbus, OH 43210, USA}
\altaffiltext{10}{Department of Physics and Astronomy, Vanderbilt
University, Nashville, TN 37235, USA; Department of Physics, Fisk
University, Nashville, TN 37208, USA}
\altaffiltext{11}{INAF - Osservatorio Astronomico di Palermo, Piazza
del Parlamento 1, 90134, Palermo, Italy}
\altaffiltext{12}{Department of Physics, University of Oxford, Keble Road, 
Oxford OX3 9UU, UK}
\altaffiltext{13}{US Naval Observatory, Flagstaff Station, P.O. Box
1149, Flagstaff, AZ 86002, USA}
\altaffiltext{14}{Astrophysics Group, Cavendish
Laboratory, J.J. Thomson Avenue, Cambridge CB3 0HE, UK}
\altaffiltext{15}{School of Physics and Astronomy, University of 
St Andrews, North Haugh, St Andrews, Fife KY16 9SS, UK}

\begin{abstract}  

Young (125 Myr), populous ($>$1000 members), and relatively nearby, 
the Pleiades has provided an anchor for stellar angular momentum
models for both younger and older stars. We used K2 to explore the
distribution of rotation periods in the Pleiades.  With more than 500
new periods for Pleiades members, we are vastly expanding the number
of Pleiads with periods, particularly at the low mass end. About 92\%
of the members  in our sample have at least one measured
spot-modulated rotation period.  For the $\sim$8\% of the members
without periods, non-astrophysical effects often dominate (saturation,
etc.), such that periodic signals might have been detectable, all
other things being equal. We now have an unusually complete view of
the rotation distribution in the Pleiades.  The relationship between
$P$ and \vmk\  follows the overall trends found in other Pleiades
studies. There is a slowly rotating sequence for
$1.1\lesssim$\vmk$\lesssim 3.7$, and a primarily rapidly rotating
population for \vmk$\gtrsim 5.0$.  There is a region in which there
seems to be a disorganized relationship between $P$ and \vmk\ for
$3.7 \lesssim$\vmk$\lesssim 5.0$. Paper II continues the discussion,
focusing on multi-period structures, and Paper III speculates about
the origin and evolution of the period distribution in the Pleiades.

\end{abstract}

\section{Introduction}
\label{sec:intro}

The three most fundamental parameters of a star are its mass, its
composition, and its angular momentum.  Together, they determine how
the star evolves from birth through the pre-main sequence phase to
main sequence hydrogen burning, and beyond, and further, whether and
how planets form and migrate. Angular momentum evolution is tied
during star formation to cloud core fragmentation processes and
stellar multiplicity, and during pre-main sequence evolution to
star-disk interactions coupled with simple radial contraction and
internal structural changes. Main sequence angular momentum evolution
is dominated by spin-down due to mass loss and core-envelope coupling
efficiencies.  Although theoretical guidance addressing these matters
for stars from Myr to Gyr ages has been significant (see,\eg, Bouvier
\etal\ 2014 and references therein), the problems to be addressed are
still lacking in empirical guidance in critical areas.

Because the Pleiades is populous (over 1000 members; e.g., Bouy \etal\
2015), relatively young (125 Myr; Stauffer \etal\ 1998a), and nearby
(136 pc; Melis \etal\ 2014), it has provided an anchor for stellar
angular momentum models for both younger and older stars. As such, we
need a thorough understanding of the rotational distribution of stars in
the Pleiades. There is ample evidence that angular momentum evolution
depends on stellar mass, so obtaining a reliable rotation distribution
for stars of a  wide range of masses is critically important.  The
NASA K2 mission (Howell \etal\ 2014), using the repurposed 1-m Kepler
spacecraft, observed the Pleiades cluster nearly continuously for 72
days, enabling us to probe rotation rates to lower masses and to
higher precision than ever before. 

The Pleiades has been extensively studied for decades (\eg, Trumpler
1921,  Hertzsprung 1947, Johnson \& Mitchell 1958), and more recent
surveys (\eg, Lodieu \etal\ 2012, Sarro \etal\ 2014, Bouy \etal\ 2015)
have identified candidate members down to at least $\sim$0.03 \msun\
(\ks$\sim$18, or $R>$22), past where K2 can obtain a viable light curve
in the Pleiades (\ks$\sim$14.5, or $K_p\sim$18). More than 1000
candidate members for the Pleiades were included in K2's Campaign 4,
down to mass $\sim$0.09 \msun. 

The rotation of stars in the Pleiades has been the subject of study
for quite some time, both spectroscopically  (\eg, Anderson, Kraft, \&
Stoeckly 1966, Stauffer \& Hartman 1987, Soderblom \etal\ 1993a,b,
Terndrup \etal\ 2000, Queloz \etal\ 1998) and photometrically (\eg,
van Leeuwen \etal\ 1987, Stauffer \& Hartmann 1987, Stauffer \etal\
1987, Prosser \etal\ 1993a,b, 1995).  There have been two recent
extensive photometric surveys of Pleiades rotation periods. Hartman
\etal\ (2010) used the Hungarian Automated Telescope Network (HATNet)
to obtain rotation periods for nearly 400 Pleiades members down to $M
\sim$0.4 \msun, with estimated completeness to $M\sim$0.7 \msun.  More
recently, Covey \etal\ (2016) present new rotation period observations
for more than 100 Pleiads from the Palomar Transient Facility (PTF),
which greatly expanded the known periods for lower mass Pleiads down
to $M\sim$0.18 \msun. These ground-based surveys, however, necessarily
were biased towards larger amplitude variability, and against periods
near $\sim$1d.

Because K2 provides precision, sensitivity, and continuous (as opposed
to diurnal) time coverage, in the present paper we push the known
periods down to lower mass and lower amplitude than has ever been done
before in the Pleiades. In the process of doing this, we have found
other repeated patterns in the light curves (LCs). We have already
scoured the K2 data for eclipsing binaries (David \etal\ 2015, 2016).
Other periods that do not appear to be spot-modulated rotation periods
are included in the Appendix. The rest of the periods are nearly all
consistent with spot-modulated rotation periods (though a few are
likely pulsation; see Paper II).

In Section~\ref{sec:obs}, we present the observations and data
reduction, as well as assembly of Pleiades members out of the 1020 K2
LCs of candidate Pleiads. The overall distribution of K2-derived
rotation rates is discussed in Section~\ref{sec:rotationdistrib}. 
Section~\ref{sec:concl} summarizes our main results. 

This is the first of three papers focused on rotation periods in the
Pleiades. Paper II, Rebull \etal\ (2016), discusses the several types
of LCs and periodogram structures that we found in the K2 data, and
some of the properties of these multi-period stars. Stauffer \etal\
(2016), Paper III, speculates about the origin and evolution of the
period distribution in the Pleiades.

\section{Observations and Methods}
\label{sec:obs}

\subsection{K2 Data}

\begin{figure}[ht]
\epsscale{0.8}
\plotone{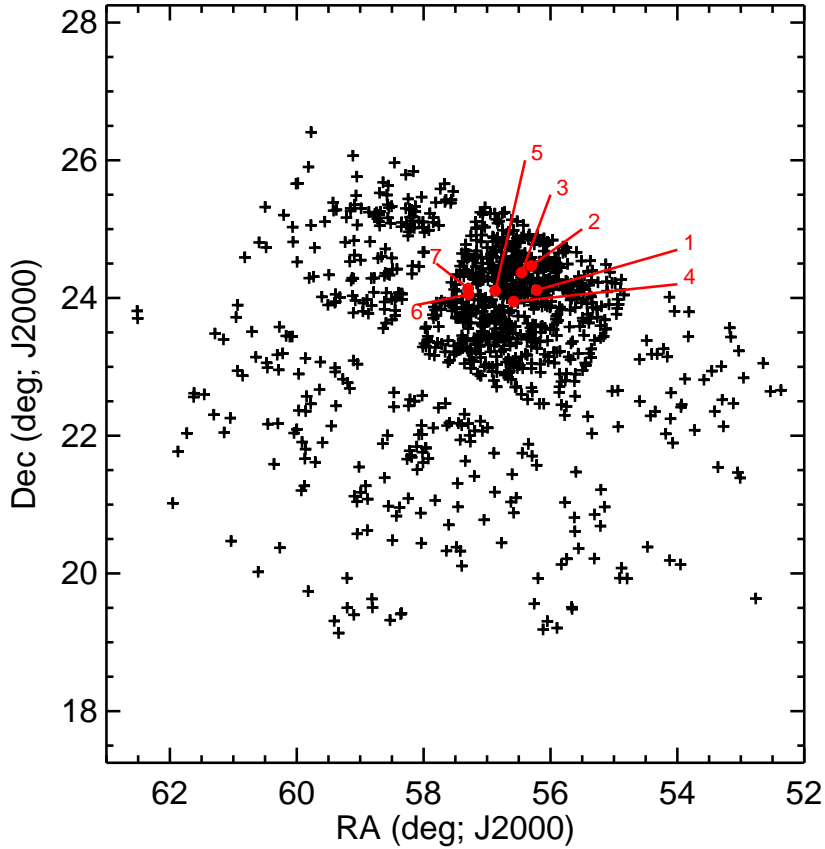}
\caption{All 1020 candidate Pleiades members with K2 LCs projected
onto the sky. Red numbers correspond as follows: 1-Electra=HII468;
2-Taygeta=HII563; 3-Maia=HII785; 4-Merope=HII980;
5-Alcyone=etaTau=HII1432; 6-Atlas=HII2168; 7-Pleione=HII2181. Note
that the entire Pleiades cluster, centered roughly on Alcyone, is not
included in the K2 fields; the tidal radius of the Pleiades is
$\sim$6$\arcdeg$. Note also the gaps between K2 detectors. }
\label{fig:where}
\end{figure}

Members of the Pleiades were observed in K2 campaign 4, which lasted
for 72d. Note that the field of view is not centered on the cluster;
see Fig.~\ref{fig:where}.  All of the stars shown were observed in the
long-cadence ($\sim$30 min exposure) mode. Thirty-four of these stars
were additionally observed in fast cadence ($\sim$1 min exposure), but
those data are beyond the scope of the present paper.  There are 1020
unique K2 long-cadence light curves.

Kepler pixel sizes are relatively large, $3.98\arcsec \times 3.98
\arcsec$, and the 95\% encircled energy diameter ranges from 3.1 to
7.5 pixels with a  median value of 4.2 pixels.  During the K2 portion
of the mission, because only two reaction wheels can be used, the
whole spacecraft slowly drifts and then repositions regularly every
0.245 d. 

We have used several different sets of LCs employing different
reductions. (1) The pre-search data conditioning (PDC) version
generated by the Kepler project and  obtained from MAST, the Mikulski
Archive for Space Telescopes. (2) A version with moving apertures
obtained following Cody \etal\ in prep. (3) A version using a
semiparametric Gaussian process model  used by Aigrain \etal\ (2015,
2016). (4) The `self-flat-fielding' approach used by Vanderburg \&
Johnson (2014) as obtained from MAST.  We removed any data points
corresponding to thruster firings and any others with bad data flags
set in the corresponding data product. The times (as shown in figures
in this and our subsequent papers) are Kepler baricentric Julian day.

We inspected LCs from each reduction approach, and we selected the
visually `best' LC from among the four, such as the LC with the least
discontinuities, or the one with the least overall trend, or the one
least subject to saturation effects, etc.  Out of our entire sample of
1020 LCs, the PDC LC was the best for $\sim$58\% of the sample, 11\%
of the LCs had the best version from Aigrain \etal, $\sim$8\% had the
best version from Cody \etal, and $\sim$5\% of the LCs were best in
the Vanderburg \& Johnson approach. It is important to note that in
most cases, the period appears as a significant peak in the
periodograms for all four LC versions, but the subtleties of the
processing mean that one version is slightly better than another and
is the one that we used to obtain the periods reported here.  In
general, the PDC LC was best for $\lesssim$3 d; both the Aigrain and
Vanderburg approaches were on average best for the longer periods. For
$\sim$18\% of the 1020, it was not clear which was the best LC, either
because the LC was saturated (too bright) or too faint, or adversely
affected by nearby bright stars, or all the LC versions were different
enough that no one LC could be selected as the best and most reliable.
None of these latter confusing LCs were found to be periodic. 

In two cases, there are pairs of lightcurves that are
indistinguishable. EPIC 211076026 and 211076042 are ADS2755A and
ADS2755B, which are sometimes jointly referred to as HII956 or HD
23479. These two stars are a visual binary with a separation of 
$\sim$0.7$\arcsec$, so close that the K2 light curves are effectively
identical. We dropped 211076026 and kept 211076042; the LC is not
periodic. EPIC 211066337 (HII298) and EPIC 211066412 (HII299) are
functionally indistinguishable light curves. They are a visual binary
separated by $\sim$6$\arcsec$. We have kept EPIC
211066337 and dropped EPIC 211066412. The net LC in EPIC 211066337 has
two periods, 6.156, and 2.932 d, and we suspect that is one period per
binary component (see Paper II).

\subsection{Finding Periods}

We looked for periodic
signals using primarily the NASA Exoplanet Archive Periodogram
Service\footnote{http://exoplanetarchive.ipac.caltech.edu/cgi-bin/Periodogram/nph-simpleupload}
(Akeson \etal\ 2013). This service provides period calculations using
Lomb-Scargle (LS; Scargle 1982), Box-fitting Least Squares (BLS;
Kov\'acs \etal\ 2002), and Plavchan (Plavchan \etal\ 2008) algorithms.
We also looked for periods using CLEAN (Roberts \etal\ 1987).  

In practice, though, the periodic signals are generally not ambigous
and any method yields very similar periods. Different LC versions can
make more of a difference in the derived period than different
period-finding algorithms because of the influence of artifacts.  We
used LS for the analysis discussed here, because most of the periodic
signals are sinusoidal.  

\begin{figure}[ht]
\epsscale{1.0}
\plotone{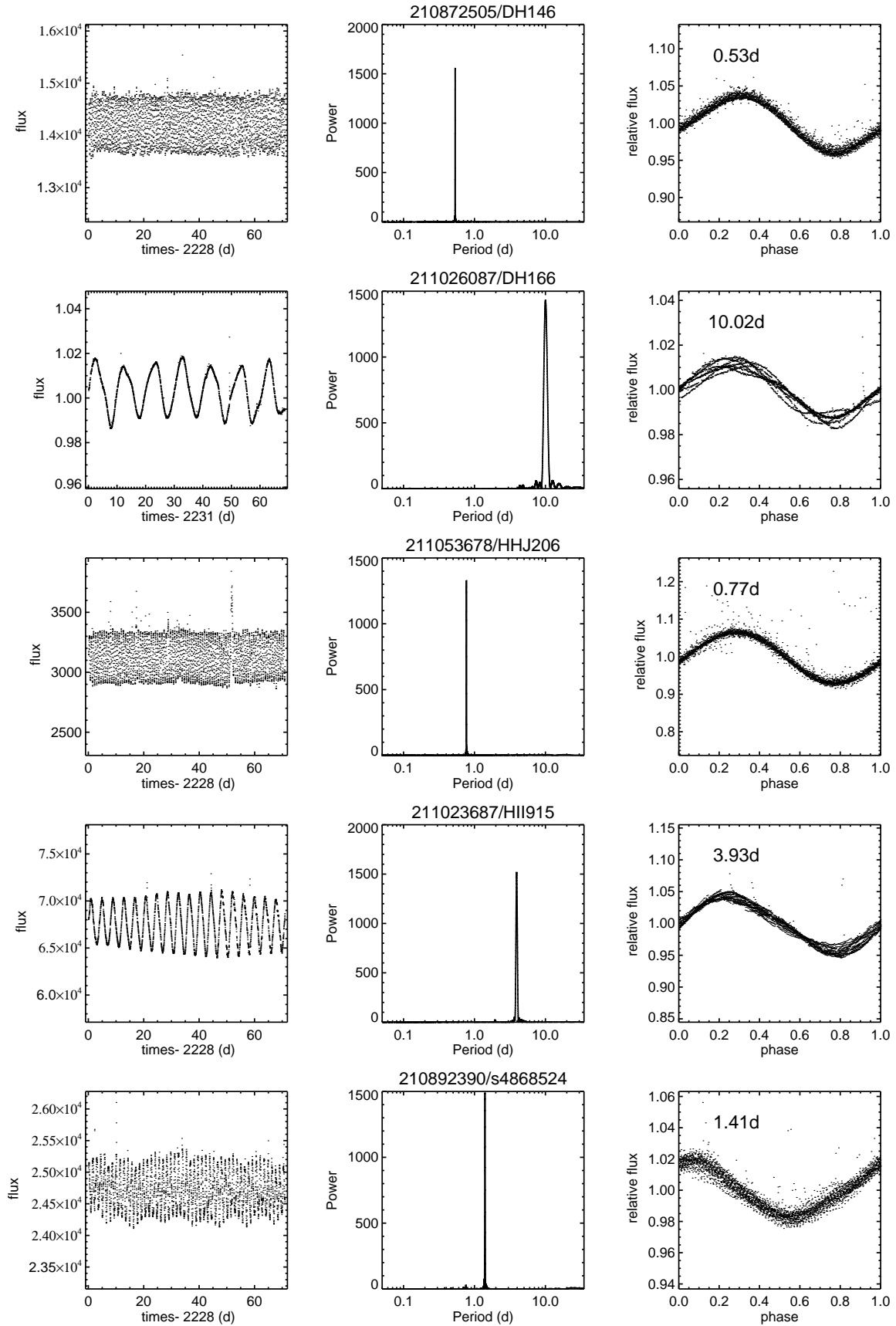}
\caption{Five examples of finding periods in the K2 Pleiades data.
Left column: full LC; middle column: LS periodogram; right column:
phased LC, with best period (in days) as indicated. Rows, in order:
EPIC 210872505/DH146, 211026087/DH166, 211053678/HHJ206,
211023687/HII915, 210892390/s4868524. These are representatives from a
range of brightnesses and periods. Note that in each case, the power
spectrum indicates unambiguously periodic signals -- the peak is so
high that little structure other than the peak  can be seen in the
power spectrum. These LCs are best interpreted as large spots or spot
groups rotating into and out of view. }
\label{fig:singlefreq}
\end{figure}

Some LCs, periodograms, and phased LCs can be found in
Fig.~\ref{fig:singlefreq}. These are representatives from a range of
brightnesses and periods. The power spectra indicate unambiguously
periodic signals -- the peak is so high that little structure other
than the peak can be seen in the power spectrum, and when there is
substructure, it is a harmonic of the main signal. (However, see Paper
II for multi-periodic stars.)  For signals like those in
Fig.~\ref{fig:singlefreq}, the false alarm probability (FAP) returned
by the LS algorithm is 0; for $\sim$97\% of the sample with periods,
the FAP of the main peak is very small, $\sim$0. For many stars, the
FAP of the second or third peak is {\em also} $\sim$0, which gives
rise to the multi-periodic discoveries in Paper II. The only
situations in which we took a star to be periodic when the FAP for the
peak calculated over the whole LC was not $\sim$0 were situations in
which, \eg,  half the LC was corrupted by instrumental effects and
thus we took a $P$ derived from the unaffected portion (which then
meant that the FAP computed for that peak on that portion of the LC
was very low), or the three stars in Sec.~\ref{sec:litP} where there
is a clear peak at the same location as others found for this star in
an independent dataset, even if the formal FAP calculated for that
peak from the K2 data was high.

For stars of the mass range considered here, the periods that we
measure are, by and large, star spot-modulated rotation periods. 
Spot modulation is the simplest explanation for sinusoidal (or
sinusoidal-like) variations where there are changes over an entire
rotation phase. 

\begin{figure}[ht]
\epsscale{1.0}
\plotone{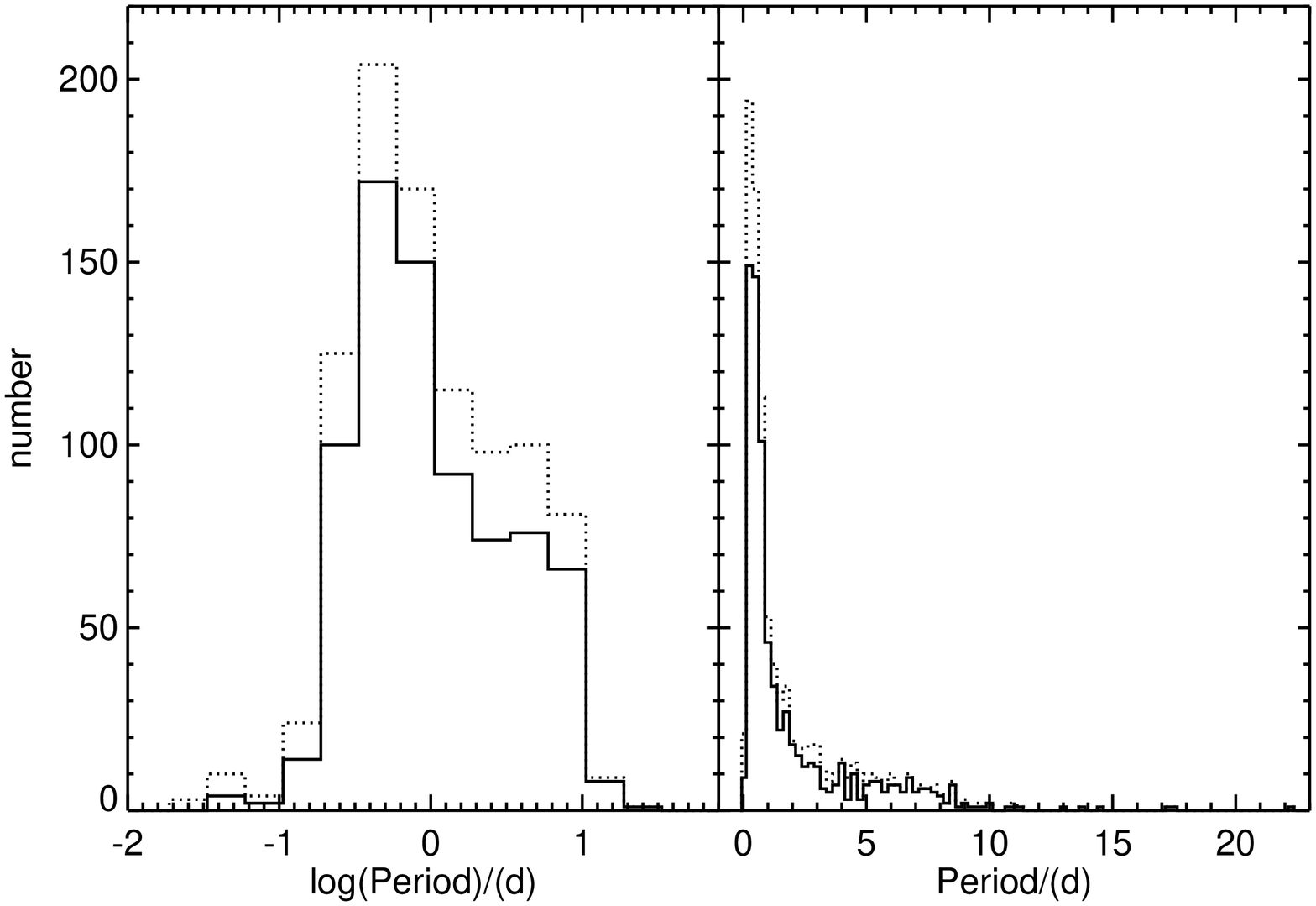}
\caption{Histograms, on the left of the log of periods, and on the
right of the linear periods, found by our analysis, in
days.  Solid line is the primary period (that which we take to be the
rotation period of the star), and dotted line is (for reference) a
histogram of all the periods found here, including the secondary,
tertiary, and quaternary periods (see Paper II).  We
limited our search to $P<35$ d, half our campaign length, but strongly
suspect that no legitimate rotation periods of Pleiades members are
$>$35d (1.54 in the log). The period distribution is strongly peaked
at $<$1d, with the maximum $P$ at 22.14d.}
\label{fig:phisto}
\end{figure}

To be conservative, we required at least 2 complete cycles of a
pattern to call it periodic, thus the maximum period we searched for
was 35 d.  We do not expect Pleiades members to be rotating more
slowly than 35d. Indeed, the distribution of periods we found (see
Figure~\ref{fig:phisto}) is strongly peaked at $<$1 day; only
$\sim$3\% of the periods over all 1020 LCs (not just members
identified in Sec.~\ref{sec:membership} below) are longer than 10 d.
Because the number of rotation periods falls off so strongly, we
suspect that few or no legitimate rotation periods of Pleiades members
are $>$35d, and our approach is not unduly biasing our derived
distribution of rotation periods in the Pleiades. There may be some
patterns that are repeated on timescales longer than 35d, but they are
not rotation periods -- the shapes of the LCs are much different than
the rotation periods in the data.

Additionally, by inspection of individual LCs, we deemed some periodic
signals with periods shorter than 35 d to not necessarily be rotation
periods. Two objects, EPIC 211082420 (HII1431) and 210822691 (AKII465)
are eclipsing binaries (see David \etal\ 2015, 2016). We have removed
these periods from our sample because they are not spot-modulated
rotation periods (AKII465 is also likely not a member of the
Pleiades). There are other eclipsing binaries in our data, but for
those, there is also a periodic signal from the primary, which we
retain here because it is likely to be a rotation rate; see, for
example, 211093684/HII2407 in David \etal\ (2015).   There are 28
additional objects that have features in their LS periodograms that
suggest possible periods $P<$35 d, but that which we believe are not
unambiguously periodic.  Those stars are listed
Appendix~\ref{app:timescales} for reference and those periods have
been removed from subsequent analysis.   

We find periods for 798 out of our sample of 1020 K2 LCs of candidate
Pleiads. However, not all of those stars may be members; see
Sec.~\ref{sec:membership}. 


\subsection{Comparison to Literature Values}
\label{sec:lit}

\subsubsection{Literature Periods}
\label{sec:litP}

\begin{figure}[ht]
\epsscale{0.8}
\plottwo{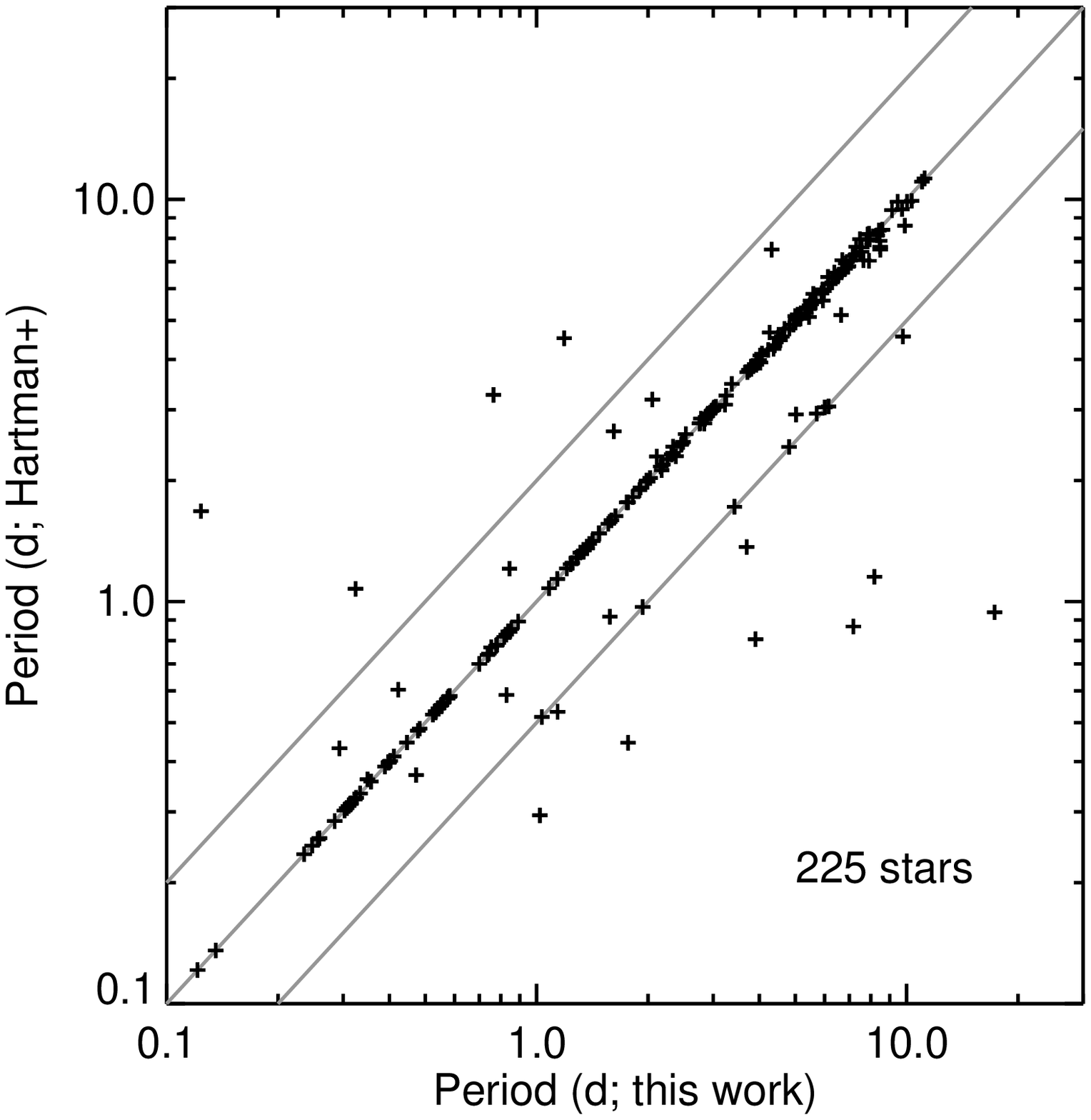}{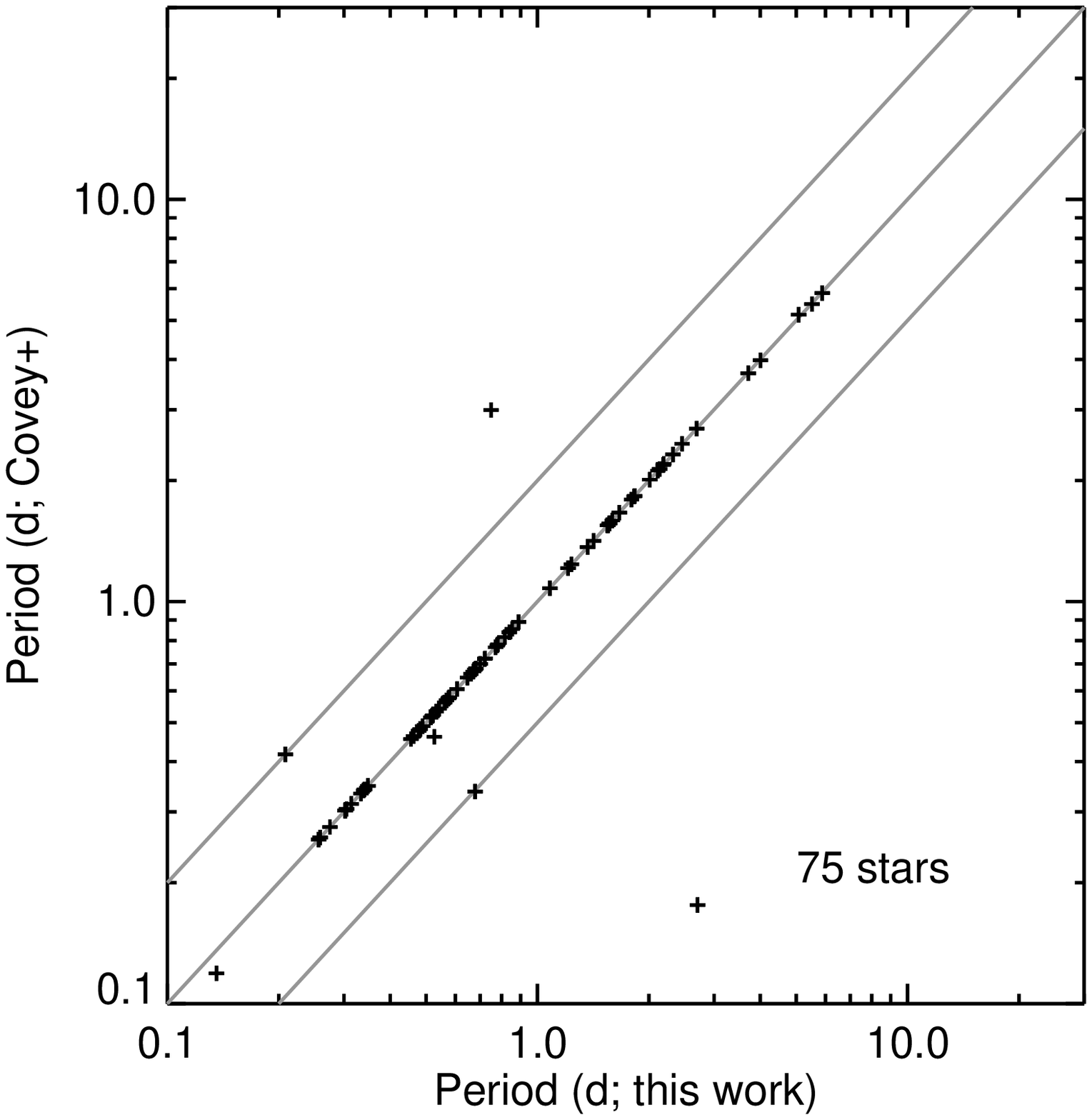}
\caption{Left: Objects with periods in both Hartman \etal\ (2010) and
this work, compared. There are 225 objects in this plot, 85\% of which
agree to 10\% or better.  Right: Objects with periods in both Covey
\etal\ (2016) and this work, compared, with those from Hartman \etal\
removed. There are 75 objects in this plot, 92\% of which agree to
10\% or better. In both cases, there are three grey lines: a 1:1
match, and the 2$P$ and $P$/2 harmonics. We
conclude that that our approach to finding periods is working at least
as well as those in the literature.}
\label{fig:comparehartmanlog}
\end{figure}

In order to verify our period-finding approach, it is useful to
compare to prior Pleiades results. There are two recent papers that
obtain periods in the Pleiades from large-field photometric
monitoring. Hartman \etal\ (2010) used HATNet and reported periods for
383 Pleiads. We have 225 periodic objects in common (given spatial and
brightness constraints), and we agree to within 10\% of the derived
$P$ for 85\% of the objects; see Figure~\ref{fig:comparehartmanlog}.
The median fractional difference ($|(P_{\rm Hartman}-P_{\rm
Rebull})|/P_{\rm Rebull}$) is 0.7\%. Covey \etal\ (2016) used PTF and
report periods for 138 Pleiads. We have 75 periodic objects in common
with this study (again, given spatial and brightness constraints), and
92\% of them agree to within 10\% of the derived $P$; see
Figure~\ref{fig:comparehartmanlog}. The median fractional difference
is 0.07\%.

For each of the targets in which we have periods that disagree (or in
which the literature reports a period that we did not find), we
inspected our LC and associated power spectrum in some detail. There
are several bright targets for which we failed to find a period where
others did; in the K2 data, the star is just too bright for the data
reduction used here.  For most of the stars where our period is very
discrepant from the published one, we believe our period is correct
for the stars at the time that we observed them. These discrepancies
may be telling us something about the long-term spot distribution
and/or spot evolution, but the details of that are beyond the scope of
the present paper. 211089068/HII1348 has a period in Hartman \etal\
that is not quite a harmonic; they report 4.562d and we have 9.773d.
In three cases, the power spectrum and phased LC derived from our data
alone are not as convincing as other sources in this study (\eg, the
FAP is not 0 for these periodogram peaks), but since they
independently recover the same period as reported in Covey \etal, we
have opted to keep them. They are 210978953/HHJ114, 211055493/JRS26,
and 211083672/HCG253.

We conclude that our approach to finding periods is working at least
as well as those in the literature.

\begin{figure}[ht]
\epsscale{0.7}
\plotone{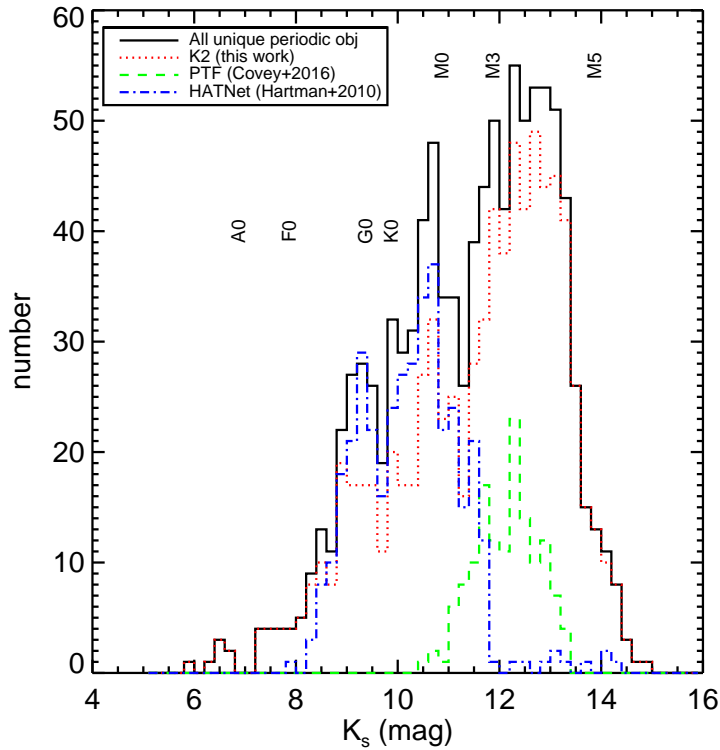}
\caption{Comparison of range of \ks\ magnitudes for the entire set of
periodic candidate Pleiads (black, solid line), with subsamples
indicated from K2 (this work), PTF (Covey \etal\ 2016), and HATNet
(Hartman \etal\ 2010) as shown. Approximate spectral types
corresponding to \ks\ in the Pleiades are annotated. There are 1184
unique objects shown here, the vast majority of which are Pleiades
members (some are not necessarily members).  The K2 study
tremendously  expands the number of known periodic objects, especially
for fainter Pleiades stars. Objects that appear in more than one study
are counted only once in the black histogram, but may appear once per
study in the colored histograms. The Hartman and Covey studies include
regions of the cluster not covered by K2. }
\label{fig:comparekrange}
\end{figure}

Figure~\ref{fig:comparekrange} demonstrates the range of \ks\
magnitudes to which the various studies are sensitive. The Hartman
\etal\ (2010) study focused on the brighter stars, and the Covey
\etal\ (2016) study focused on the fainter stars. This work, with K2,
increases the numbers of periods known overall, but makes a more
significant contribution of new periods for the fainter (lower mass)
stars. Note that this plot includes periods for the 798 out of our 
1020 K2 light curves; the literature reports periods for $\sim$500
(candidate) Pleiads, so we have more than doubled the number of known
periods for candidate Pleiads. (However, not all of the K2 LCs are for
likely member stars; see Sec.~\ref{sec:membership}.)

In the subsequent analysis here, we made a decision to not include 
periods from these literature studies for objects that did not have a
K2 light curve, which omits $\sim$220 periods (not all of which may be
members). Since the K2 target selection is primarily biased in
position (Fig.~\ref{fig:where}), and since rotation period is not a
function of the location in the cluster, this does not affect our
conclusions. 


\subsubsection{Literature \vsini}

\begin{figure}[ht]
\epsscale{0.6}
\plotone{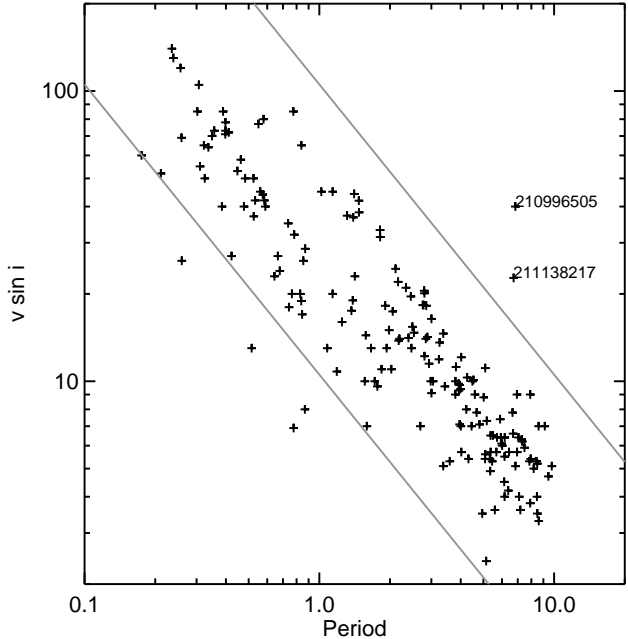}
\caption{$P$ (in days) vs.~\vsini\ (in km s$^{-1}$) for stars in this
study for which there are \vsini\ values in the literature. The grey
lines correspond to the expected relationship between $P$ and $v \sin
i$ (=$(2\pi R \sin i)/P)$ for $i=$90$\arcdeg $and 6$\arcdeg$,  
assuming $R$=0.5$R_{\odot}$. The $P$
and \vsini\ agree well, except for two cases -- both of which are
earlier type stars with a secondary component, where the \vsini\ is
probably from the primary, and the $P$ is from the lower-mass
secondary. They are 210996505/HII1132, and 211138217/HII1766. }
\label{fig:pvsini}
\end{figure}

Much early work on rotation in the Pleiades was done on projected
rotational velocities, \vsini.  Figure~\ref{fig:pvsini} shows the
relationship between $P$ and \vsini\ for stars in this study for which
there are \vsini\ values in the literature (see
section~\ref{sec:suppdata}). The $P$ and \vsini\ agree well overall,
which is an indication that we are measuring the rotation rate for
these stars. The $P$ and \vsini\ do not agree well for two cases,
210996505/HII1132, and 211138217/HII1766. These are both earlier type
stars which likely have a secondary component, where the \vsini\ is
probably from the primary, and the $P$ is from the lower-mass
secondary (see Paper III).

Both Jackson \& Jeffries (2010) and Hartman \etal\ (2010) have already
looked at the distribution of $\sin i$ in the Pleiades in detail. This
kind of analysis is limited by the number of \vsini\ values known;
although we are adding many periods here, there are no new \vsini\
values.

\subsection{Supporting Data from the Literature}
\label{sec:suppdata}

We assembled a catalog of photometric data for all of our targets from
the literature, including   Johnson \& Mitchell (1958), Stauffer
\etal\ (1998a,b), Stauffer \etal\ (2007),  Kamai \etal\ (2014), and Bouy
\etal\ (2015). We added to this  data from the Two-Micron All Sky
Survey (2MASS; Skrutskie \etal\ 2006), from the Spitzer Space
Telescope (Werner \etal\ 2004), including measurements from Sierchio
\etal\  (2010) and the Spitzer Enhanced Imaging Products,
SEIP\footnote{http://irsa.ipac.caltech.edu/data/SPITZER/Enhanced/SEIP/overview.html},
from the Widefield Infrared Survey Explorer (WISE; Wright \etal\
2010), from SIMBAD's listing of the Tycho catalog (ESA 1997), and from
the United States Naval Observatory (USNO) Robotic Astrometric
Telescope (URAT; Zacharias \etal\ 2015).  

Ideally, we would have \teff\ or mass for all of our targets. However,
those quantities can be very model-dependent.  Because we preferred to
keep our discussion of the new K2 rotation period data on an empirical
basis to the extent possible, our goal was to use an observed color as
the proxy for mass or \teff.  The broad-band color that acts as the
best such proxy over the entire mass range for which we have periods
is \vmk.  While \ks\ is widely available from 2MASS, $V$ is harder to
find. We only have measured $V$ band photometry for about half of the
periodic stars; it was necessary to estimate $V$ magnitudes from other
photometry for the rest.   

\begin{figure}[ht]
\epsscale{0.5}
\plotone{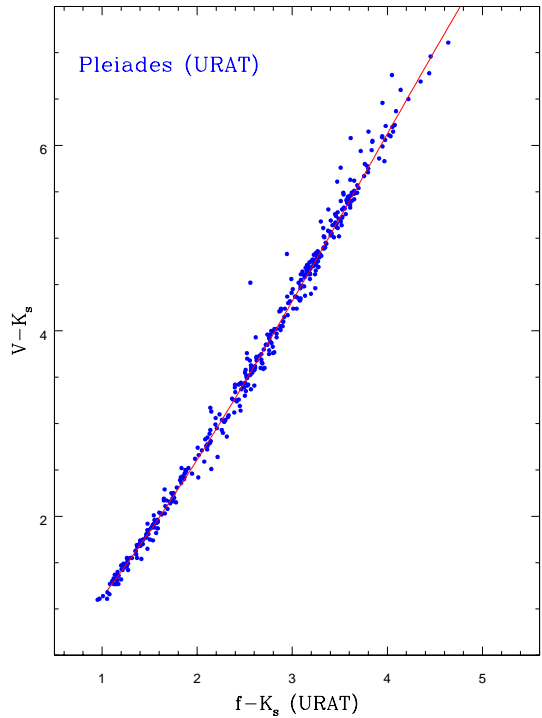}
\caption{Empirical relationship between $V-K_s$ and $f-K_s$ (where the
$f$ comes from URAT). The best-fit line (equation 3) is the magenta
line. We used this relationship to obtain estimates of \vmk\ for those
stars for which we had no $V$ measure; see text.  }
\label{fig:uratrelation}
\end{figure}

The highest quality $V$ band photometry we have is from phototube
photometry reported in Johnson \& Mitchell (1958), Landolt (1979),
Stauffer \& Hartmann (1987) or references therein, or CCD photometry
from Kamai \etal\ (2014).  Additional $V$ band photometry,
generally for fainter members, was obtained using CCD cameras on small
telescopes by Prosser \etal\ (1991) and Stauffer \etal\ (1998b).   For
the remaining stars (mostly faint M dwarfs), we have adopted measured
photometry at bands near in wavelength to $V$. Specifically, we have
adopted $g$ or $r$ magnitudes from SDSS-filter images reported in Bouy
\etal\ (2013) or Bouy \etal\ (2015), or ``$f$" magnitudes (a very
broad-band red filter) provided with the initial release of the URAT
catalog (Zacharias \etal\ 2015) for all Pleiades members for which
those quantitities are reported.   For the stars for which we also
have measured $V$ magnitudes, we have then derived  transformations
between $g-K_s$, $r-K_s$ and $f-K_s$ and $V-K_s$;
Figure~\ref{fig:uratrelation} shows the data for one such
transformation.   For each of these three data sources, the photometry
appears to have similar accuracies as the available $V$ band
photometry, and the transformations are well-defined and not strongly
curved.  The three polynomial relations are:
\begin{equation}
V-K_s = 0.3837 + 0.48719\times (g-K_s) + 0.08564\times (g-K_s)^2 - 0.00488\times (g-K_s)^3
\hspace{0.5cm} \rm{for}\hspace{0.25cm}  1.75 < g-K_s < 7.75
\end{equation}
\begin{equation}
V-K_s = -0.2991 + 1.47462\times (r-K_s) - 0.07522\times (r-K_s)^2 +
0.00394\times (r-K_s)^3  \hspace{0.5cm} \rm{for}\hspace{0.25cm} 1.0 < r-K_s < 6.25
\end{equation}
\begin{equation}
V-K_s = -0.004 + 0.91784\times (f-K_s) + 0.23683\times (f-K_s)^2 - 0.02080\times (f-K_s)^3
\hspace{0.5cm} \rm{for}\hspace{0.25cm} 1.0 < f-K_s < 4.75
\end{equation}

For most stars, we have these estimated $V-K_s$ values from all three
sources.  When we have a measured $V-K_s$, we use that; when we do not
have a measured $V-K_s$, we use the average estimated $V-K_s$. Note
that this is \ks, not $K_p$, that is K-short from 2MASS, not Kepler
magnitude; \ks\ is used throughout this paper, and not $K_p$.

We assume that the typical reddening in the direction of the Pleiades
applies:  $A_v$ = 0.12,  $A_K$ = 0.01, $E(B-V)$ = 0.04 (Crawford \&
Perry 1976). There are four
stars with K2 LCs that have larger reddening (HII476, HII870, HII1039,
\& HII1136); for these, we used reddening corrections from Soderblom
\etal\  (1993b) and Breger (1986). Below, Tables~\ref{tab:bigperiods},
\ref{tab:bigaperiodic}, and \ref{tab:nonmembers} include the observed
$V$ and \ks\ where available, from which one can derive the observed
$V-K_s$; the \vmk\ that we used is explicitly included.

\subsection{Membership and Definition of Sample}
\label{sec:membership}

In order to establish the best possible set of Pleiades members, we
evaluated each object using a combination of proper motions and
photometric position in an optical color-magnitude diagram (CMD).  We
primarily used membership probabilities based on  recent proper motion
studies, Bouy \etal\ (2015); see also Sarro \etal\ (2014) and Lodieu
\etal\ (2012)\footnote{All objects analyzed by Lodieu \etal\ appear in
Bouy \etal. Some Lodieu \etal\ members are reassigned in Bouy \etal\
which has better proper motions and photometry. See discussion in
Sarro \etal\ (2014).}. For objects where the membership probability
and the photometric position were inconsistent, we evaluated stars on
a case-by-case basis, comparing information from many sources, such as
positions and proper motions, radial velocities, X-ray flux, IR flux,
and H$\alpha$ equivalent width.  These values are from the literature,
including 
Trumpler (1921), 
Hertzsprung (1947), 
Johnson \& Mitchell (1958), 
Ahmad \etal\ (1965),
Iriarte (1967),  
Artyukhina and Kalinina (1970),
Jones (1970, 1973), 
Breger (1972, 1986), 
Morel \& Magnenat (1978),
Landolt (1979), 
Vasilevskis \etal\ (1979),  
Stauffer \etal\ (1984), 
van Leeuwen \etal\ (1986, 1987), 
Stauffer \& Hartmann (1987), 
Jameson \& Skillen (1989)
Mermilliod \etal\ (1997, 2009),  
Micela \etal\ (1990, 1999), 
Prosser \etal\ (1991), 
Stauffer \etal\ (1991), 
Rosvick \etal\ (1992), 
Soderblom \etal\ (1993a), 
Kazarovets (1993), 
Hodgkin \etal\ (1995), 
Schilbach \etal\ (1995), 
Martin \etal\ (1996), 
Wang \etal\ (1996), 
Burkhart \& Coupry (1997), 
Zapatero Osorio \etal\ (1997),
Belikov \etal\ (1998), 
Stauffer \etal\ (1998a,b), 
Queloz \etal\ (1998), 
Malaroda \etal\ (2000),
Pinfield \etal\ (2000, 2003), 
Ducati (2002),
Deacon \& Hambly (2004), Li \etal\ (2004), 
Scholz \& Eisl\"offel (2004), 
Mermilliod (2006),
Fox Machado \etal\ (2006, 2011), 
Gebran \& Monier (2008),
Renson \& Manfroid (2009),
Roeser \etal\ (2010; PPMXL), 
Lodieu \etal\ (2012),  
Zacharias \etal\ (2013; UCAC4), 
Cottaar \etal\ (2014), and
Zacharias \etal\ (2015; URAT). 

\begin{figure}[ht]
\epsscale{0.8}
\plotone{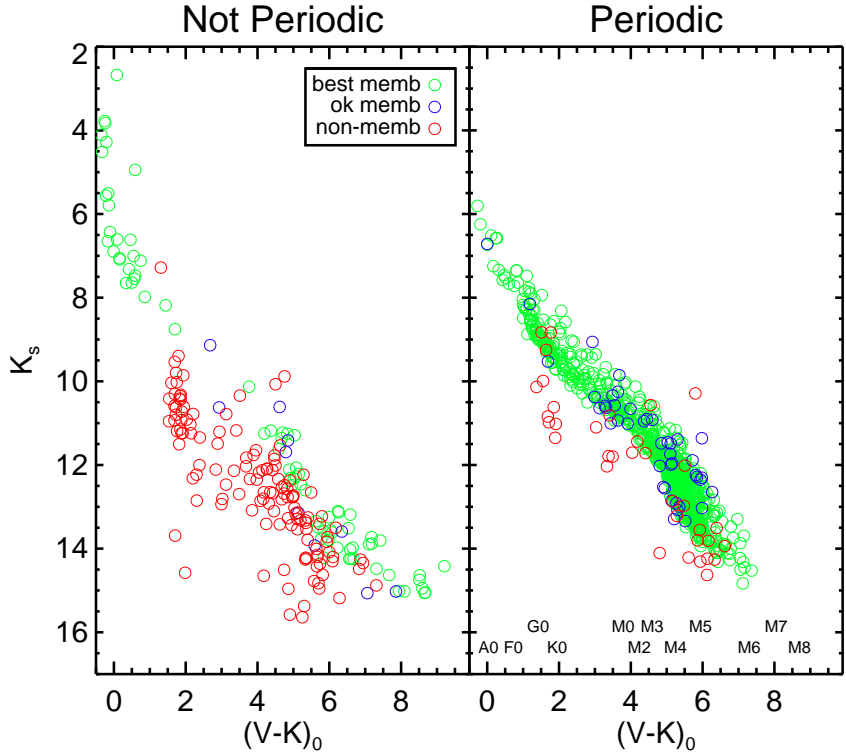}
\caption{Optical CMD (\ks\ vs.\ \vmk) for stars with K2 LCs and for
which we could calculate \vmk.  Left panel is stars not measured to be
periodic, and right panel is stars for which we could measure periods.
Spectral types for a given \vmk\ are as shown in the bottom of the
right panel. Green symbols are our best, highest-confidence sample of
members, blue symbols correspond to those lower-confidence members (ok
members), and red symbols are non-members. Most of the members have
periods, and most are comfortably in the expected location of the main
sequence. The NM have considerably more scatter and fewer periods. }
\label{fig:optcmd}
\end{figure}

Our membership analysis usually began with the location of the star in
a $V$ vs.\ $V-K$ CMD; in many cases, we also looked at other CMDs in
order to make sure that a bad measurement in one band was not causing
a discrepant CMD location.   Next usually we looked at proper motion
measurements in multiple studies (including the all-sky surveys like
PPMX, UCAC4, URAT, etc.), again in order to attempt to minimize the
influence of single ``bad" measurement.  If those steps did not yield
an unambiguous decision, we next looked at all references to the star
in SIMBAD in order to, for example, determine if any previous study
had determined radial velocities or lithium equivalent widths or other
data from which membership could be inferred (such as X-ray data).  In
a very few cases, we obtained new spectra to help determine membership
(see Paper III).  This process was qualitative in the sense that we
weighted all of the information in a subjective manner.  However, the
process was also extensive, with each star considered individually and
with all available information considered in detail.  Based on the
location of these stars in the CMD (Fig.~\ref{fig:optcmd}) and the
fact that most of the non-members show no period in their K2 data (also
Fig.~\ref{fig:optcmd}), we believe in the great majority of cases we
have made the right decision.

As a result of this analysis, we have a set of our highest confidence
members, for which there is considerable data supporting membership
(often abbreviated as `best members'), and a set of non-members (NM).
There is also a set of lower-confidence members (often abbreviated as
`ok members'), where the evidence for membership is suggestive but not
conclusive  (\eg, all the proper motion studies said it was an
unambiguous member, but it was slightly too high or too low in one of
the optical CMDs, and had insufficient data to place it in the other
optical CMDs). Our final list of members (best or ok) is in
Table~\ref{tab:bigperiods} (for the periodic members) and
Table~\ref{tab:bigaperiodic} for the rest, and is what we carry
forward here. The list of objects we investigated with K2 LCs but that
we believe are not Pleiades members appears in the Appendix, along
with derived periods where relevant.  Figure~\ref{fig:optcmd} shows
the optical CMD, \ks\ vs.\ \vmk, for stars with K2 LCs and for which
we could obtain or calculate \ks\ and  \vmk. Many of the objects we
took to be NM are clearly in a position inconsistent with membership. 

Figure~\ref{fig:optcmd} also shows our effective bright and faint
cut-offs. For $K_{\rm s}\lesssim 6$ and $K_{\rm s}\gtrsim 14.5$ (or
$M_K\lesssim 0.5$ and $M_K\gtrsim 9$), the K2 LCs are either
too bright or too faint to yield reliable periods using our approach.
These objects are dropped from our sample going forward, and appear as
a list in the Appendix.

\begin{deluxetable}{p{2cm}cp{7cm}}
\tabletypesize{\scriptsize}
\floattable
\rotate
\tablecaption{Star Counts
\label{tab:lccounts}}
\tablewidth{0pt}
\tablehead{\colhead{Name} & \colhead{Number} & \colhead{Description} }
\startdata
Initial sample	&	1020	&	All K2 LCs of candidate
Pleiades members. 	 \\ 
Best members	& 775 &	Highest-confidence (our determination)
Pleiades members with K2 light curves, and neither too bright nor too
faint (6$<$\ks$<$14.5).   	\\ 
OK members	& 51 &	Lower-confidence (our determination) Pleiades
members with K2 light curves, and neither too bright nor too faint
(6$<$\ks$<$14.5).   	\\ 
The sample, a.k.a. the sample of members	& 826 &	The set of all
high-confidence (`best') plus lower-confidence (`ok') members that
are  neither too bright nor too faint (6$<$\ks$<$14.5); a.k.a.
`members of the right brightness range.'	 \\ 
The periodic sample &	759	&	The subset of all
high-confidence (`best') plus lower-confidence (`ok') members that
are  neither too bright nor too faint (6$<$\ks$<$14.5) and are found
to be periodic by us in these K2 data. 		\\ 
\enddata
\end{deluxetable}

Our set of members consists of 799 high-confidence (`best') Pleiades
members and 54 more lower-confidence (`ok') members, for a total of
853.  Thus, we find that 167 of the candidate Pleiads with K2 LCs are
unlikely to be members (see Appendix). Omitting the too bright and too
faint stars for our sample, there are 775 high-confidence members,
with 51 more lower-confidence members (for a total of 826 members).
Out of those 775 (best members), 716 (92.4\%) have at least one
measured period that we believe in the overwhelming majority of cases
to be a rotation period and due to star spots.   Including the lower
confidence members, 759/826 (91.9\%) have at least one measured period
that we believe to be the rotation period. Table~\ref{tab:lccounts}
summarizes the most important of these numbers.
Table~\ref{tab:bigperiods} includes all of these members and their
measured periods. This sample of members (both `best' and `ok') that
are within 6$<$\ks$<$14.5 is hereafter the set of `members of the
right brightness range,' and is what our analysis is based on (unless
otherwise specified). An online-only figure set with one set of plots
(like those in Fig.~\ref{fig:singlefreq}) for each star can be found
in Appendix~\ref{sec:efig}.

We also scoured the literature for any information about binarity.
This information came from:
Abt \etal\ (1965), 
Anderson \etal\ (1966),
Stauffer \etal\ (1984), 
Liu \etal\ (1991),
Mermilliod \etal\ (1992), 
Rosvick \& Mermilliod (1992), 
Soderblom \etal\ (1993b), 
Bouvier \etal\ (1997), 
Queloz \etal\ (1998), 
Raboud \& Mermilliod (1998), 
Geissler \etal\ (2012), and 
Kamai \etal\ (2014).  
We note here that most of these literature surveys focused on the
brighter sources, and there are K2 data for many fainter stars. We
discuss more about binaries below, primarily in
Sec.~\ref{sec:binariesmain}, and in Papers II and III. (Note that
Paper III also includes a description of how we identified photometric
binaries.)

\subsection{Members Not Detected As Periodic}
\label{sec:aperiodicmembers}

As can be seen in Fig.~\ref{fig:optcmd}, about 8\% of the sample are 
not detected as periodic in our data (see Appendix~\ref{app:ndap} for
example LCs and power spectra). For these stars, one or more of these
criteria are met: (a) no periodogram peaks with very low FAP in the LS
output; (b) periodogram peak(s) change position significantly
between LC versions, or the purported periodic signal appears as a
peak in the periodograms in only one LC version; (c) phased LC 
does not look convincing (\eg, a wide distribution of fluxes at most
phases) because the pattern is not well-repeated from cycle to cycle;
(d) LC obviously and signficantly affected by instrumental effects
(\eg, bimodal distribution of flux values originating from saturated
pixels); (e) rarely, the repeated pattern is not consistent with spot
modulation. A list of those in the last category appear in
Appendix~\ref{app:timescales}.

Some of these not-detected-as-periodic members can be found at nearly
every color. The K2 data are exquisite, and we expect all stars to
rotate, and low-mass stars as young as the Pleiades should have large
starspots. There are several possible explanations as to why we do not
detect these stars to be periodic. These stars could have periods much
longer than 35d, which is very unlikely for the Pleiades. Despite our
best efforts, these stars could actually be non-members (and thus
could have a period much longer than 35 d; with only 72 d of data, it
would be hard to reliably identify a period much longer than 35 d). 
It could be that these stars have periodic variations on timescales
$<$35 d but at a lower level than we can detect, perhaps from smaller
spots/spot groups. The stars could have a rotation axis that is
pole-on, such that there really is little to no variation detectable
from our Solar System.  Alternatively, they could have disorganized
spots distributed more or less homogeneously that preclude a reliably
periodic signal in the LC. In about half of the cases, however, these
LCs are corrupted by instrumental effects in the data reductions we
have; either the stars are too bright themselves for reliable LCs, or
nearby bright stars adversely affect the extracted LCs\footnote{Other
investigators have methodologies to extract photometry and derive
periods for very bright (saturated) stars (\eg, White \etal\ 2015),
but these are not included here, in part because we do not have access
to those data reductions, but also because we are primarily interested
in the FGKM stars.}. There are also several LCs that are just
effectively too faint (poor signal-to-noise ratio, SNR) for viable
periods to be extracted from the data reductions we have. A period
might have been detected for many of these not detected as periodic
member stars if the saturation level was higher, or the exposures
different (longer for the poor SNR, shorter for the saturated), or if
the star was located elsewhere on the CCD.  Since we already detect
periods in a very large fraction of the members, were it not for these
non-astrophysical effects, the fraction could be even closer to 1.
This is different than prior studies of rotation in clusters, and it
means that we have an unusually complete view of the rotation
distribution in the Pleiades.

Basic parameters for these stars not detected as periodic are listed
in Table~\ref{tab:bigaperiodic}. Five of these stars have reported
periods in the literature which we do not recover; see
Table~\ref{tab:bigaperiodic}.

\clearpage

\begin{deluxetable}{cl}
\tabletypesize{\scriptsize}
\floattable
\tablecaption{Contents of Table: Periods and Supporting Data for Periodic Pleiades
Members\label{tab:bigperiods}}
\tablewidth{0pt}
\tablehead{\colhead{Label} & \colhead{Contents}}
\startdata
EPIC & Number in the Ecliptic Plane Input Catalog (EPIC) for K2\\
coord & Right ascension and declination (J2000) for target \\
Vmag & V magnitude (in Vega mags), if observed\\
Kmag & \ks\ magnitude (in Vega mags), if observed\\
vmk0 & \vmk\ -- dereddened $V-K_s$, directly observed (if $V$ and \ks\
exist) or inferred (see text)\\
P1 & Primary period, in days (taken to be rotation period)\\
P2 & Secondary period, in days\\
P3 & Tertiary period, in days\\
P4 & Quaternary period, in days\\
ampl & Amplitude, in magnitudes, of the 10th to the 90th percentile\\
LC & LC used as `best'\tablenotemark{a}\\
memb & Membership indicator: Best, OK, or NM \\
Plit & Literature (rotation) period, in days, if available\\
vsini & Literature $v \sin i$, in km s$^{-1}$, if available \\
\enddata
\tablenotetext{a}{LC1=PDC, from MAST; LC2=version following Cody
\etal\ in prep; LC3=version following Aigrain \etal\ (2015, 2016);
LC4=version reduced by Vanderburg \& Johnson (2014) and downloaded
from MAST.}
\end{deluxetable}

\begin{deluxetable}{ccccccc}
\tabletypesize{\scriptsize}
\tablecaption{Supporting Data for Pleiades
Members Not Detected to be Periodic in the K2
Data\tablenotemark{a}\tablenotemark{b}\label{tab:bigaperiodic}}
\tablewidth{0pt}
\tablehead{\colhead{EPIC} & \colhead{RA, Dec (J2000)}
& \colhead{other name} & \colhead{$V$ (mag) }
& \colhead{\ks\ (mag) }  & \colhead{\vmk\tablenotemark{c} (mag) }
& \colhead{Membership} 
}
\startdata
210784603 & 033103.57+193805.1 & s3289407 & \ldots &   10.13 &    3.77 &  best  \\
210899735 & 033202.35+212310.8 & \ldots & \ldots &   12.11 &    4.89 &  best  \\
210904850 & 033211.53+212756.1 & UGCSJ033211.55+212755.7 & \ldots &   13.90 &    7.11 &  best  \\
210971138 & 033310.49+223119.3 & DH027 &   15.55 &   11.25 &    4.19 &  best  \\
211029507 & 033518.74+232621.0 & DH045 & \ldots &   13.98 &    6.22 &  best  \\
\enddata
\tablenotetext{a}{This table is available in its entirety in the online
version. A portion is shown here to demonstrate its form and content.}
\tablenotetext{b}{Five stars from this table have periods in the
literature: EPIC 210931896 (4.25 d), 211036390 (9.92 d), 211083301
(3.70 d), 211056483 (1.565 d), and 210917230 (9.46 d). We do not
recover these periods from the K2 data.}
\tablenotetext{c}{Dereddened $V-K_s$, directly observed (if $V$ and \ks\
exist) or inferred (see text).}
\end{deluxetable}


\section{Period and Period-Color Distributions}
\label{sec:rotationdistrib}

As discussed above, the overwhelming majority of the periods we have
determined are spot-modulated rotation periods of the stars. We can
now proceed to investigate the distribution of rotation rates.  In
this section, we investigate the rotation distribution against \vmk\
color.  Note that that we have selected only one $P$ (and \vmk) to be
representative of the rotation period (and color) in the $\sim$22\% of
the stars for which there are more than one period recovered (see
Paper II).

\subsection{Morphology of $P$ vs. \vmk}

\begin{figure}[ht]
\epsscale{0.8}
\plotone{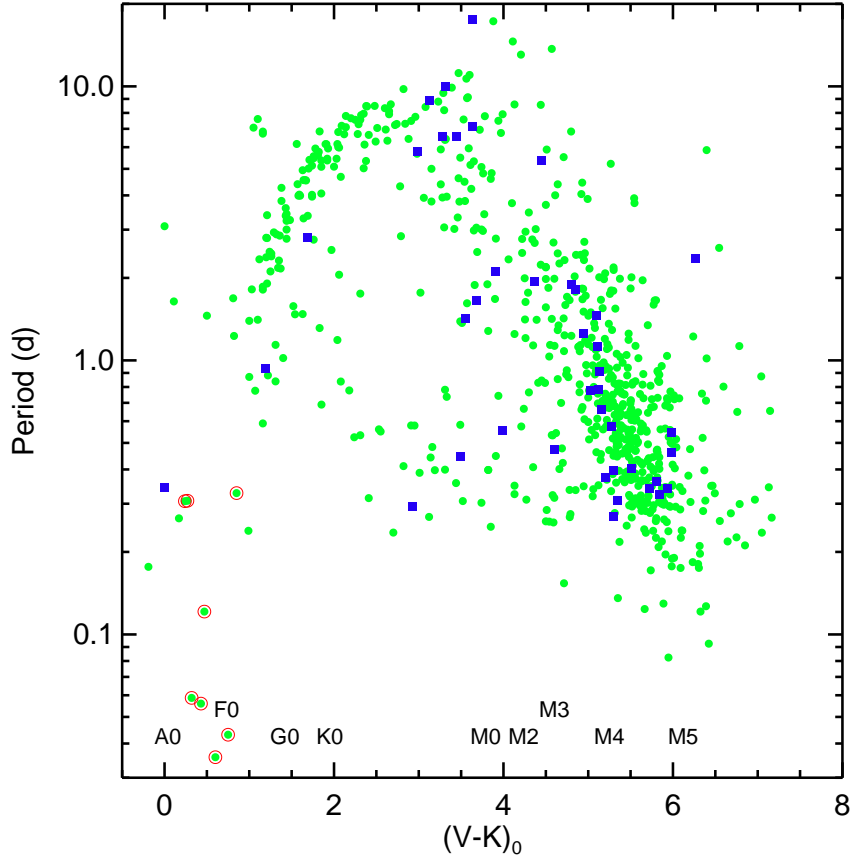}
\caption{Plot of $P$ vs.~\vmk\ for the best members (green dots) and
the lower confidence members (blue squares). Pulsators ($\delta$
Scutis from Paper II) have an additional red circle.  The
distribution follows the same overall trends found in other Pleiades
studies. There is a slowly rotating main sequence for
$1.1\lesssim$\vmk$\lesssim 3.7$, and a primarily rapidly rotating
population for \vmk$\gtrsim 5.0$. }
\label{fig:pvmk}
\end{figure}

Figure~\ref{fig:pvmk} shows the relationship between $P$ and \vmk\ for
the sample. It follows the overall trends found in other Pleiades
studies (\eg, Covey \etal\ 2016, Hartman \etal\ 2010). There is a
slowly rotating sequence for $1.1\lesssim$\vmk$\lesssim 3.7$
(2$\lesssim P \lesssim$11 d), and a primarily rapidly rotating
population for \vmk$\gtrsim 5.0$ (0.1 $\lesssim P \lesssim$ 2 d). 
There is a region in which there seems to be a disorganized
relationship between $P$ and \vmk\ between $3.7\lesssim$\vmk$\lesssim
5.0$  (0.2$\lesssim P \lesssim$ 15 d). 

Another important thing to note is that the lower-confidence members
still follow the overall trends here; there is no compelling evidence
from this plot per se to move those lower-confidence members into the
non-member set. (In contrast, see Appendix~\ref{app:nm} and
Fig.~\ref{fig:pvmkNM} below.)

Among the long-period outliers in this plot, there are five stars with
periods longer than 12 days, one of which is a lower-confidence
member, but four of which are high confidence members. Those long-$P$
outliers are curious, since they seem out of place relative to the
other members. These stars are discussed further in Paper III, though
we highlight one here. We have taken EPIC 210855272/DH668  to have a
period of 17.6 d for the reasons discussed in Paper II; that seems to
be the best period. However, if we take $P$=8.9 d (the other peak that
appears in the periodogram), then this star would no longer be a
long-period outlier; for its \vmk, it would have a $P$ more consistent
with other stars of its color. It has H$\alpha$ in absorption, but
this may be acceptable for \vmk=3.6 mag.  At that color, it has the
largest H$\alpha$ absorption equivalent width in the Pleiades,
comparable to a field star at that color. It has a 60\% chance of
being a member according to Deacon \& Hambly (2004), and a 72\% chance
of being a member in Bouy \etal\ (2015); based on that, we have it as
a lower-confidence member. 

Among the blue short-period outliers, the stars we identify as
pulsators in Paper II have demonstrably shorter periods on average
(and are among the bluest stars) than the rest of the ensemble, and
this matches expectations. There are other very blue stars that are
not quite among the shortest periods. Their LC morphology are more
suggestive of rotation than pulsation.  The stars with  $P\sim$0.3 d
could also be pulsators (see Paper II).   Normal A and F stars should
not have spots, though Am stars could have spots (\eg, Balona \etal\
2015). They could also be unresolved binaries, where the \vmk\
corresponds to the primary, but the $P$ corresponds to the fainter,
lower-mass, spotted secondary.

\subsection{Binaries}
\label{sec:binariesmain}

Whether a star is single or is one component of a binary could affect
the measured rotation period we detect in the K2 Pleiades data.  This
might be true because the formation mechanism for single and binary
stars might have different dependencies on either the initial angular
momentum of the collapsing cloud core or on how much of that angular
momentum is retained in the ZAMS descendents of that process.
In addition, being a member of a binary will affect the photometric
colors we measure and the signal-to-noise properties of any periodic
signature that we measure in the K2 data.   We therefore have searched 
for the possible influence of binarity on the period distribution that 
we have measured.
The discussion
about the influence of binarity continues in Papers II and III.

\begin{figure}[ht]
\epsscale{0.8}
\plotone{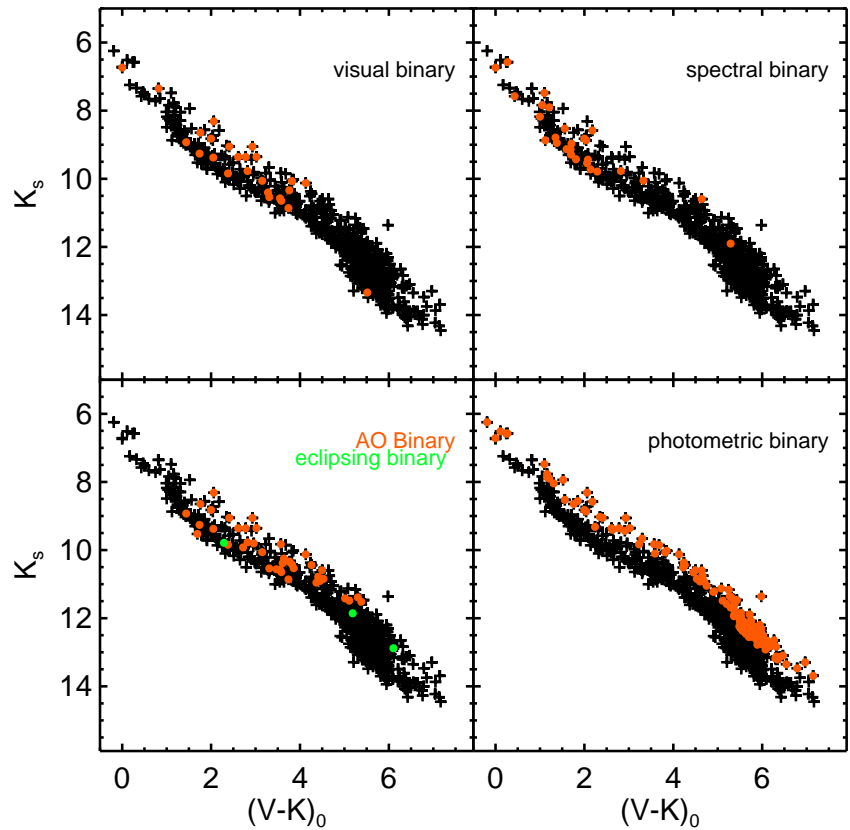}
\caption{Plot of \ks\ vs.~\vmk, highlighting the binaries listed in
the literature as found by a variety of literature methods (top left:
visual binaries; top right: spectroscopic binaries; bottom left: AO
[orange] and eclipsing binaries [green]), and in the lower right, the
binaries just from position in the CMD as used here. The many
approaches to identifying binaries identify different stars as binary.
}
\label{fig:optcmdbinary}
\end{figure}

\begin{figure}[ht]
\epsscale{0.8}
\plotone{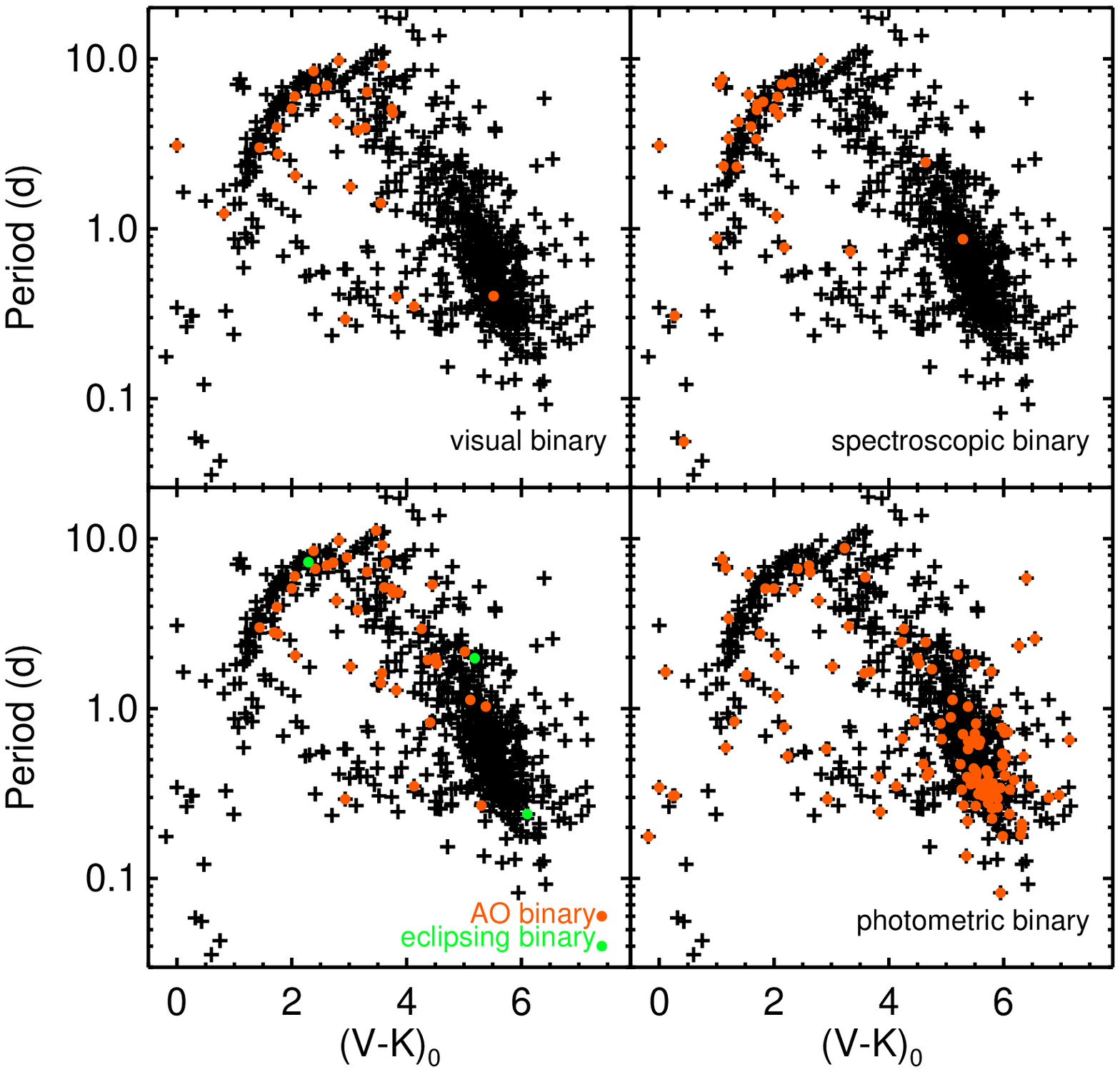}
\caption{Plot of $P$ vs.~\vmk, highlighting the binaries listed in the
literature as found by a variety of literature methods (top left:
visual binaries; top right: spectroscopic binaries; bottom left: AO
[adaptive optics; orange] and eclipsing binaries [green]), and in the
lower right, the binaries just from position in the CMD used here
(Fig.~\ref{fig:optcmd}). While the many approaches to identifying
binaries identify different stars as binary, there is no clear and
obvious trend in the overall $P$ vs.~\vmk\ that simply segregates
binaries from single stars; see text.  }
\label{fig:pvmk5binary}
\end{figure}

We assembled binary information from the literature (see
Sec.~\ref{sec:membership}) from spectroscopy, radial velocities, high
spatial resolution imaging, and new Robo-AO data (Riddle \etal\ in
prep). The advantage of this inhomogenous data set is that we have a
chance of identifying all of the binaries, but the disadvantage is
that many existing surveys were limited spatially or limited by
stellar brightness such that lack of binarity from this collection of
data may actually reflect a lack of information rather than anything
else. Figure ~\ref{fig:optcmdbinary} shows where these literature
binaries fall in the CMD; Figure~\ref{fig:pvmk5binary} shows $P$
vs.~\vmk\ with the literature binaries highlighted. In these figures,
visual binaries, spectroscopic binaries, binaries from adaptive optics
(AO) observations, and eclipsing binaries are indicated separately. It
is clear that most of the literature methods focused on bluer
(brighter) stars. About 10\% of the sample is tagged binary in the
literature.

We can also use the optical CMD assembled here (Fig.~\ref{fig:optcmd})
to more uniformly identify photometric binaries by the stars' location
in the CMD (see Paper III for details of this process). The advantage
of this approach is that we can identify binaries with uniform
sensitivity through the whole viable range of our data. The
disadvantage is that we will miss binaries whose masses are
significantly different from each other (i.e., causing only small
shifts in the CMD). Figure ~\ref{fig:optcmdbinary} shows the CMD, and
Figure~\ref{fig:pvmk5binary} shows $P$ vs.~\vmk, for photometric
binaries selected via this approach; about 16\% of the sample is
tagged binary.

The slow sequence has relatively few binaries identified, except for
the spectroscopic binaries, and these are nearly all cases where no
secondary has been directly detected. In these cases, the secondary is
generally much fainter than the primary, and the secondary is
therefore unlikely to be significantly affecting the K2 light curves.
Even in the cases of the visual binaries in the slow sequence, many of
the secondaries are much fainter than their primaries, and here too,
the secondaries are unlikely to have much of an impact on the K2 light
curves of these stars. Since the primary stars are in the `right
place' in the $P$ vs.~\vmk\ diagram for stars of their \vmk\ (mass),
we infer that there is little influence on the primary's rotation rate
by the much lower mass secondary for these binaries.

We continue the discussion about the influence of binarity in Papers
II and III.

\subsection{Amplitudes}
\label{sec:amplitudes}

\begin{figure}[ht]
\epsscale{0.6}
\plotone{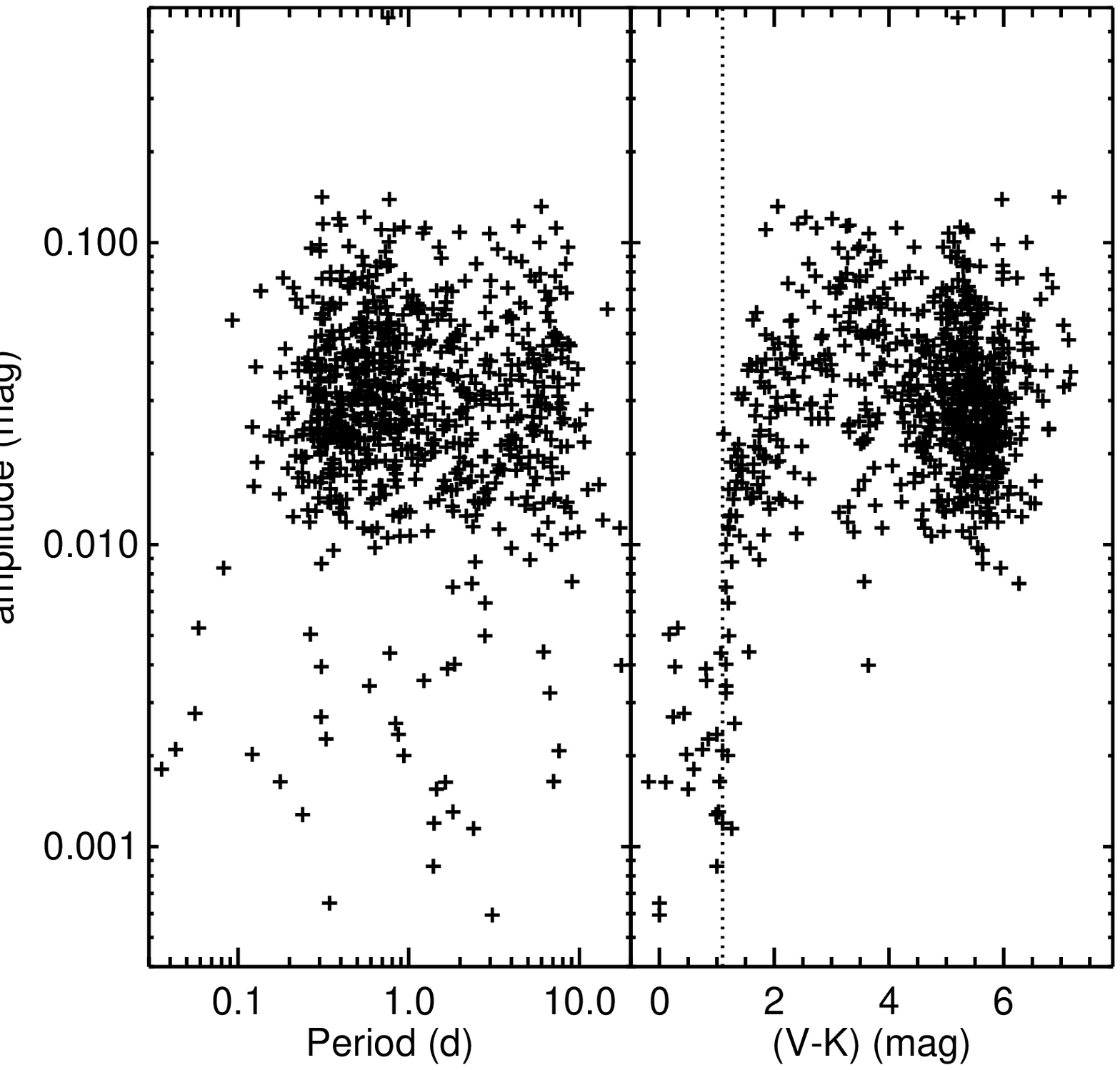}
\caption{The amplitude (from the 10th to the 90th percentile), in
magnitudes, of the periodic light curves, against $P$ and \vmk. 
The vertical dotted line is at \vmk=1.1 (see
Paper III, where there is a linear version of this plot). 
Stars bluer than about \vmk$\sim$1.1 have clearly lower amplitudes. 
The median of the amplitude distribution (with or without the
\vmk$<$1.1 stars) is 0.030 mag.  }
\label{fig:ampldistrib}
\end{figure}

We calculated the amplitude of the light curves in magnitudes by
assembling the distribution of all points in the light curve, taking
the log of the 90th percentile flux, subtracting from that the log of
the 10th percentile flux, and multiplying by 2.5.
Figure~\ref{fig:ampldistrib} plots that amplitude against both $P$ and
\vmk\ for the periodic light curves. While the lower amplitude
variations are found at all periods, they tend to cluster at bluer
colors; stars bluer than about \vmk$\sim$1.1 (see Paper III) have
distinctly lower amplitudes. Some of these stars are likely pulsators,
which accounts for the lower amplitude. Some are probably unresolved
binaries, where the amplitude of the flux variation from the companion
(causing the periodicity) is lessened by the flux from the primary. 
The median of the amplitude distribution (with or without the
\vmk$<$1.1 stars) is 0.030 mag.

The outlier with the very large amplitude is
211010517/UGCSJ040234.77+230828.4 and it has a large amplitude because
of a large-scale trend that is superimposed on the periodic light
curve. We have taken it to be one of the best (high-quality) members,
but it is just barely in the expected location in the CMD to be placed
in the best member subset.

Aside from the outliers at the small and large amplitude ends of the
distribution, there does not seem to be a trend with color or period.

\section{Conclusions}
\label{sec:concl}

We have presented the first part of our analysis of the K2 Pleiades
lightcurves, in the process vastly expanding the number of Pleiades
members known with periods, particularly at the low mass end. About
92\% of the observed Pleiades members have at least one measured
period, the overwhelming majority of which we believe to be
spot-modulated rotation periods.  For the $\sim$8\% of the members
without periods, non-astrophysical effects often dominate (saturation,
etc.), such that periodic signals might have been detectable, all
other things being equal. We now have an unusually complete view of
the rotation distribution in the Pleiades. 

The overall relationship between $P$ and \vmk\  follows the overall
trends found in other Pleiades studies. There is a slowly rotating
sequence for $1.1\lesssim$\vmk$\lesssim 3.7$ (2$\lesssim P \lesssim$11
d), and a primarily rapidly rotating population for \vmk$\gtrsim 5.0$
(0.1 $\lesssim P \lesssim$ 2 d).  There is a region in which there
seems to be a disorganized relationship between $P$ and \vmk\ between
$3.7\lesssim$\vmk$\lesssim 5.0$ (0.2$\lesssim P \lesssim$ 15 d).

Thanks in no small part to the many low-mass fast rotators, the
distribution of periods peaks strongly at $<$1 day; only $\sim$3\% of
the periods are longer than 10 d. The typical amplitude of the
variation (between 10-90\% of the distribution of points) is
$\sim$0.03 mag. Some much lower amplitudes can be found at the bluest
colors, which could be from pulsation or a consequence of binarity
(where the lower-mass, fainter star is responsible for the
spot-modulated rotation period). 

Our periods agree well with the literature periods and literature
\vsini. There is no simple way to distinguish binaries from single
stars in the $P$ vs.\ \vmk\ parameter space.

About 70\% of the periodic stars have a single, essentially stable
period. However, we have discovered complicated multi-period behavior
in Pleiades stars using these K2 data, and we discuss this further in
Paper II.  Paper III (Stauffer \etal\ 2016) continues the discussion
by speculating on the origin and evolution of the periods in the
Pleiades.

\clearpage

\acknowledgments

We thank R.~Stern and T.~David for helpful comments on draft
manuscripts. ACC acknowledges support from STFC grant ST/M001296/1.

Some of the data presented in this paper were obtained from the
Mikulski Archive for Space Telescopes (MAST). Support for MAST for
non-HST data is provided by the NASA Office of Space Science via grant
NNX09AF08G and by other grants and contracts. This paper includes data
collected by the Kepler mission. Funding for the Kepler mission is
provided by the NASA Science Mission directorate. 

This research has made use of the NASA/IPAC Infrared Science Archive
(IRSA), which is operated by the Jet Propulsion Laboratory, California
Institute of Technology, under contract with the National Aeronautics
and Space Administration.    This research has made use of NASA's
Astrophysics Data System (ADS) Abstract Service, and of the SIMBAD
database, operated at CDS, Strasbourg, France.  This research has made
use of data products from the Two Micron All-Sky Survey (2MASS), which
is a joint project of the University of Massachusetts and the Infrared
Processing and Analysis Center, funded by the National Aeronautics and
Space Administration and the National Science Foundation. The 2MASS
data are served by the NASA/IPAC Infrared Science Archive, which is
operated by the Jet Propulsion Laboratory, California Institute of
Technology, under contract with the National Aeronautics and Space
Administration. This publication makes use of data products from the
Wide-field Infrared Survey Explorer, which is a joint project of the
University of California, Los Angeles, and the Jet Propulsion
Laboratory/California Institute of Technology, funded by the National
Aeronautics and Space Administration. 

\facility{Kepler} \facility{K2} \facility{Spitzer}
\facility{2MASS} 
facilty{IRSA} facility{NASA Exoplanet Archive}
facility{Simbad} facility{Vizier}

\appendix

\section{Examples of Members Not Detected as Periodic}
\label{app:ndap}

Section~\ref{sec:aperiodicmembers} above mentions member stars not
detected by us as periodic. Here are 6 examples of these kinds of
stars. In the top row, 211029507/DH045 and 210804032/DH354 both have
flat LCs with no significant periodogram peaks. In the second row, 
210967607/DH335 and  210998086/PELS174 both have LCs compromised by
saturated pixels, either from themselves or a nearby bright star. In
the last row,  210784603/s3289407 and  210837336/PELS063 both have
somewhat of a repeated pattern, but this pattern is irregular enough
that we have designated these as having a `timescale', not a rotation
period; see \S\ref{app:timescales}).

\begin{figure}[ht]
\epsscale{1.0}
\plotone{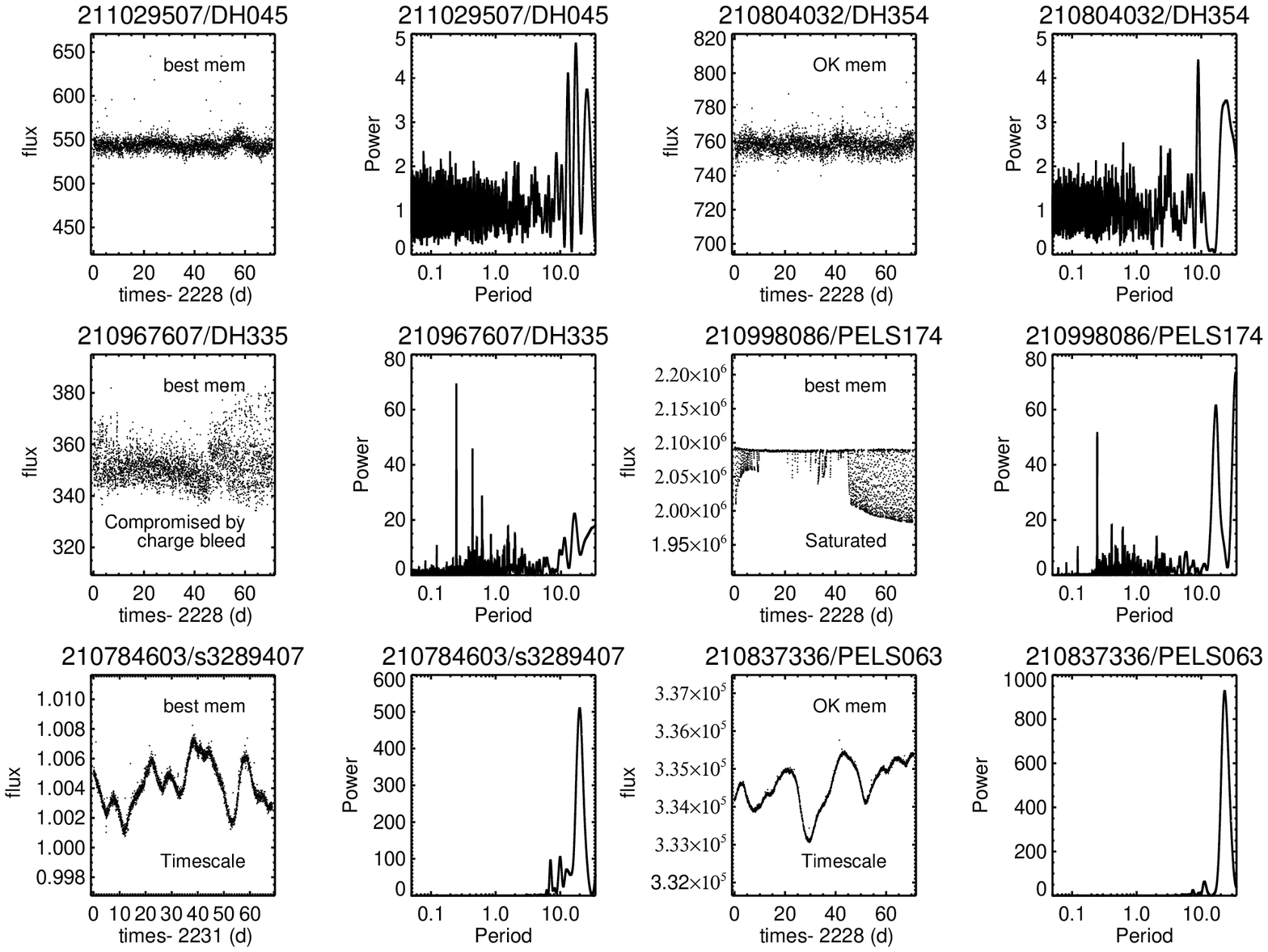}
\caption{Full LC and power spectrum for 6 stars not detected as
periodic by us. Stars,
in order, L to R, top to bottom:
211029507/DH045 (best member; no significant periodogram peaks),
210804032/DH354 (OK member; no significant periodogram peaks),
210967607/DH335 (best member, but compromised by charge bleed),
210998086/PELS174 (best member, but saturated),
210784603/s3289407 (best member, but pattern irregular enough that
this is a `timescale', not a rotation period; see
\S\ref{app:timescales}),
210837336/PELS063 (OK member, but pattern irregular enough that
this is a `timescale', not a rotation period; see
\S\ref{app:timescales}).}
\label{fig:npcallouts}
\end{figure}

\section{Timescales}
\label{app:timescales}

\begin{figure}[ht]
\epsscale{0.5}
\plotone{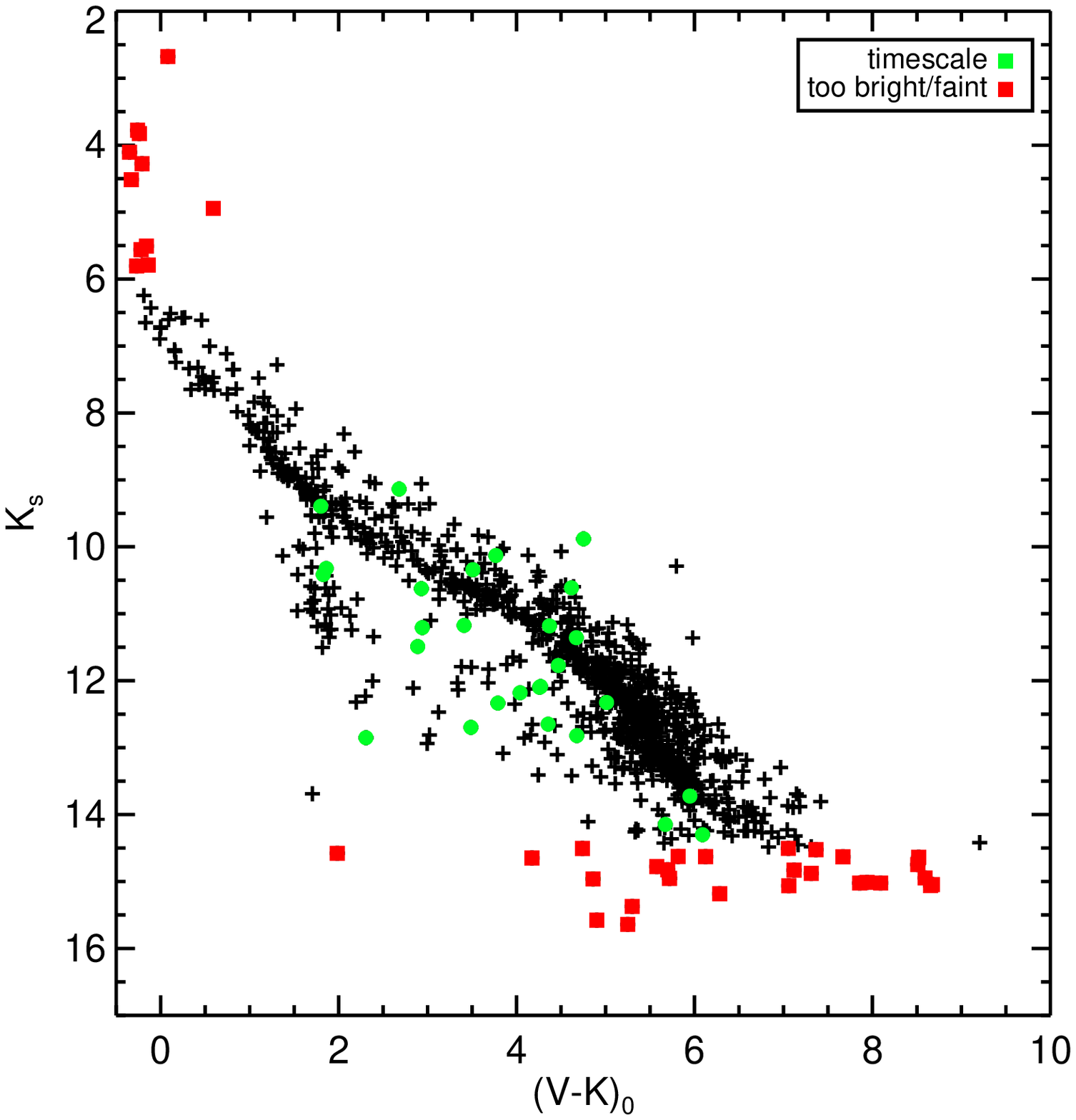}
\caption{Optical CMD (\ks\ vs.~\vmk) for the
ensemble (+), those sources with timescales as opposed to spot periods
(green dot), and those sources that are too bright or faint for us to
obtain a complete sample of periods (red square). Timescales only
appear for $2<$\vmk$<8$. It is not clear if these signatures come
from spots or not. }
\label{fig:optcmdts}
\end{figure}

For some objects, we found a period during our analysis, but
individual inspection of the light curves suggests that whatever is
causing the repeating pattern is not a spot-modulated rotation period.
We have opted to describe this as a `timescale'  rather than a period.
In several cases, they are also non-members. They do not really have a
preferential color; see Fig.~\ref{fig:optcmdts}. However, this Figure
also demonstrates that many of the timescale objects are 
photometric non-members. For comparison, Fig.~\ref{fig:npcallouts}
includes two examples of  timescale objects that are members.

\begin{figure}[ht]
\epsscale{1.0}
\plotone{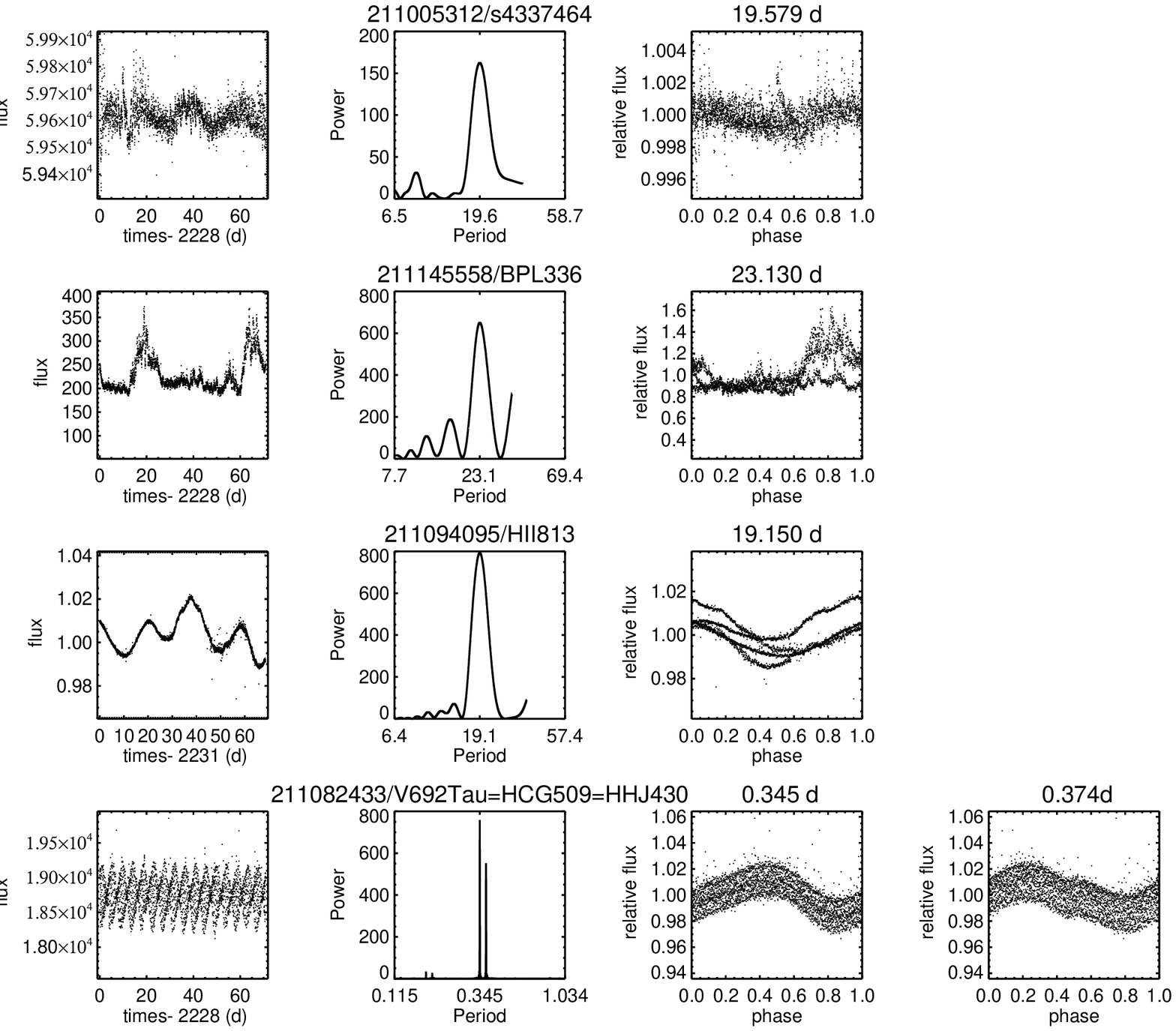}
\caption{Full LC (first column), power spectrum (second column) and
phased LC for primary period (third column) and secondary period
(fourth column) for apparent non-members (rows, in order)
211005312/s4337464 (most likely of these to be a spot-modulated
rotation period, but a NM), 211145558/BPL336 (resembles the very young
`bursters' found in NGC 2264),  211094095/HII813 (could plausibly be a
spot period, but a NM), and 211082433/V692Tau=HCG509=HHJ430 (LC that
looks like other members, but it is too high in the optical CMD to be
a member).}
\label{fig:nmcallouts}
\end{figure}

The repeating pattern in EPIC 211005312/s4337464 is the most
borderline case, by which we mean that this LC could conceivably be a
spot-modulated signal; see Fig.~\ref{fig:nmcallouts}. However, it is a
NM.  

In one case, EPIC 211145558=BPL336, the LC texture resembles the `bursts'
from NGC 2264, characterized by Stauffer \etal\ (2014) as accretion bursts
(see Fig.~\ref{fig:nmcallouts}). This object is not a member of the
Pleiades; it could be a member of Taurus, suggesting that an interpretation
of accretion bursts is not impossible. It does not have a clear IR excess
(it is detected only at the shortest two bands of WISE, and
[3.4]$-$[4.6]=0.3 mag).  It is located very far to the East and somewhat to
the North of the Pleiades cluster center, towards where Taurus members are.
Another possibility of course is that all the structure is instrumental,
not real, but this structure is found in all of the LC versions we have. 
It is listed as a photometric member in Pinfield \etal\ (2003), though the
proper motions in URAT are not consistent with Pleiades membership, and
Bouy \etal\ (2015) also have it as a clear NM.

For completeness, we note that the actual eclipses in the eclipsing
binaries, which also have a repeating pattern that is not a rotation
period, are not included here, but appear in David \etal\ (2015,
2016).

\begin{deluxetable}{cccccp{4cm}}
\tabletypesize{\scriptsize}
\tablecaption{Timescales\label{tab:timescales}}
\tablewidth{0pt}
\tablehead{\colhead{EPIC} & \colhead{RA, Dec (J2000)} &\colhead{Other name}
&\colhead{Timescale (d)} & \colhead{Membership} &\colhead{Notes} }
\startdata
210784603	&	033103.57+193805.1	&	s3289407	&	20	&	best memb	&	\ldots	\\
210909681	&	033326.45+213229.8	&	s4679029	&	22	&	NM	&	\ldots	\\
211018285	&	033652.52+231545.2	&	DH065	&	16.81	&	ok memb	&	Possibly rotation period or possibly timescale.	\\
211071563	&	033935.45+240706.3	&	HHJ407	&	33	&	best memb	&	\ldots	\\
211063756	&	034218.86+235922.2	&	SK687	&	18	&	NM	&	\ldots	\\
211020453	&	034512.69+231746.5	&	HII580	&	$\sim$25	&	NM	&	\ldots	\\
211114664	&	034554.25+244806.7	&	WCZ114	&	$\sim$18	&	NM	&	\ldots	\\
210837725	&	034705.34+202639.5	&	s4798986	&	$\sim$30	&	ok memb	&	\ldots	\\
211045908	&	035122.47+234219.1	&	BPL234	&	$\sim$26	&	NM	&	\ldots	\\
211026586	&	035146.55+232334.7	&	BPL241	&	24	&	NM	&	\ldots	\\
210914077	&	035202.47+213637.7	&	DH735	&	$\sim$28	&	NM	&	\ldots	\\
211020039	&	035203.61+231721.5	&	SK202	&	29	&	NM	&	\ldots	\\
210837336	&	035208.80+202618.9	&	PELS063	&	23	&	ok memb	&	\ldots	\\
210928539	&	035225.91+215031.8	&	DH752	&	30	&	NM	&	\ldots	\\
211132831	&	035242.50+250702.9	&	SRS33701	&	$\sim$12.5	&	NM	&	\ldots	\\
211063255	&	035321.59+235854.2	&	SK132	&	$\sim$35	&	NM	&	\ldots	\\
211044588	&	035359.92+234100.3	&	BPL296	&	$\sim$5.2	&	NM	&	\ldots	\\
211082538	&	035507.08+241725.8	&	SK67	&	$\sim$25-30	&	NM	&	\ldots	\\
211051964	&	035549.89+234822.7	&	BPL332	&	$\sim$35	&	NM	&	\ldots	\\
211145558	&	035621.72+252110.4	&	BPL336	&	23	&	NM	&	``bursts, '' not spots. 	\\
211050613	&	035626.19+234703.5	&	DH835	&	15-30	&	NM	&	Hard to know which period is right.	\\
210751596	&	035721.70+190803.4	&	UGCSJ035721.71+190803.2	&	$\sim$25	&	NM	&	only see repeated pattern in one LC version.	\\
211110418	&	035852.12+244348.5	&	UGCSJ035852.13+244348.3	&	$\sim$25	&	NM	&	\ldots	\\
211137552	&	040049.31+251210.9	&	DH895	&	$\sim$29	&	NM	&	\ldots	\\
211038138	&	040105.79+233444.3	&	UGCSJ040105.81+233443.8	&	$\sim$20	&	best memb	&	\ldots	\\
211005312	&	040154.08+230331.6	&	s4337464	&	19.579	&	NM	&	This could be a spot period. 	\\
211026906	&	040437.17+232351.5	&	s5092529	&	$\sim$25	&	best memb	&	\ldots	\\
211045801	&	041001.97+234212.3	&	s4634206	&	long, $\sim$50d?	&	NM	&	\ldots	\\
\enddata
\end{deluxetable}

\section{Stars that are too bright or too faint for this sample}
\label{app:brightfaint}

We empirically determined that our brightness and faintness limits are
effectively $K_{\rm s}\lesssim 6$ and $K_{\rm s}\gtrsim
14.5$, respectively. Sometimes, despite these limits, we were still
able to derive a period (and in two cases, two distinct periods).
Here, we list those objects that are too bright or too faint for our
sample.

\begin{deluxetable}{ccccl}
\tabletypesize{\scriptsize}
\tablecaption{Targets that are too bright or too faint.\label{tab:brightfaint}}
\tablewidth{0pt}
\tablehead{\colhead{EPIC} & \colhead{RA, Dec (J2000)} &\colhead{Other name}
&\colhead{Period(s) (d)} &\colhead{membership} }
\startdata
210967871	&	033228.28+222806.5	&	DH019	&	\ldots	&	ok memb	\\
211075945	&	034133.76+241118.6	&	BPL24	&	0.496079	&	NM	\\
211068507	&	034248.16+240401.4	&	BPL45	&	\ldots	&	best memb	\\
211110493	&	034300.16+244352.5	&	BPL49	&	\ldots	&	best memb	\\
211096282	&	034340.29+243011.4	&	BPL62	&	\ldots	&	best memb	\\
211097372	&	034353.54+243111.6	&	BPL66	&	\ldots	&	best memb	\\
211082490	&	034448.21+241722.2	&	Caeleno=HII447	&	\ldots	&	best memb	\\
200007769	&	034452.53+240647.8	&	Electra=HII468	&	\ldots	&	best memb	\\
211116936	&	034509.73+245021.3	&	HII541	&	0.674473, 0.647851	&	best memb	\\
200007772	&	034512.50+242802.1	&	Taygeta=HII563	&	\ldots	&	best memb	\\
200007771	&	034549.60+242203.7	&	Maia=HII785	&	\ldots	&	best memb	\\
210976082	&	034550.40+223606.0	&	BPL78	&	\ldots	&	ok memb	\\
211073549	&	034550.64+240903.7	&	BPL79	&	\ldots	&	best memb	\\
211099592	&	034554.47+243316.2	&	Asterope=HII817	&	\ldots	&	best memb	\\
200007770	&	034619.58+235654.1	&	Merope=HII980	&	\ldots	&	best memb	\\
211086019	&	034623.11+242036.3	&	BPL100	&	\ldots	&	ok memb	\\
211138940	&	034702.53+251345.6	&	BPL123	&	\ldots	&	NM	\\
211106625	&	034705.70+244003.7	&	BPL124	&	0.267422	&	best memb	\\
211094511	&	034712.09+242832.0	&	BPL132	&	\ldots	&	best memb	\\
211088076	&	034717.91+242231.6	&	BPL137	&	\ldots	&	NM	\\
210983090	&	034723.97+224237.3	&	BPL142	&	0.304403, 0.247935	&	best memb	\\
200007767	&	034729.08+240618.4	&Alcyone=etaTau=HII1432	&	\ldots	&	best memb	\\
211102808	&	034739.00+243622.3	&	BPL152	&	\ldots	&	best memb	\\
211090981	&	034819.00+242512.9	&	BPL172	&	\ldots	&	NM	\\
211028385	&	034820.81+232516.5	&	HII1823	&	\ldots	&	best memb	\\
210993392	&	034825.60+225212.3	&	BPL179	&	\ldots	&	NM	\\
200007768	&	034909.74+240312.1	&	Atlas=HII2168	&	\ldots	&	best memb	\\
200007773	&	034911.21+240812.0	&	Pleione=HII2181	&	\ldots	&	best memb	\\
211048942	&	035125.55+234521.3	&	BRB14	&	\ldots	&	best memb	\\
211029838	&	035144.91+232639.5	&	BPL240	&	\ldots	&	best memb	\\
211123454	&	035157.11+245706.5	&	DH729	&	\ldots	&	NM	\\
210914077	&	035202.47+213637.7	&	DH735	&	\ldots	&	NM	\\
211131711	&	035323.34+250550.5	&	BPL284	&	\ldots	&	NM	\\
211069099	&	035334.54+240438.2	&	BPL287	&	\ldots	&	NM	\\
211135437	&	035415.27+250952.3	&	BPL306	&	\ldots	&	best memb	\\
211140205	&	035444.19+251511.1	&	BPL316	&	\ldots	&	NM	\\
211115616	&	035523.07+244905.2	&	BPL327	&	0.231512	&	best memb	\\
211103918	&	035622.33+243723.4	&	DH832	&	\ldots	&	NM	\\
211080216	&	035634.19+241513.1	&	DH838	&	\ldots	&	NM	\\
210981512	&	035639.65+224112.2	&	DH842	&	\ldots	&	NM	\\
211085541	&	035758.48+242008.8	&	DH861	&	\ldots	&	NM	\\ 
\enddata
\end{deluxetable}

\clearpage

\section{Non-Members}
\label{app:nm}

\begin{figure}[ht]
\epsscale{0.5}
\plotone{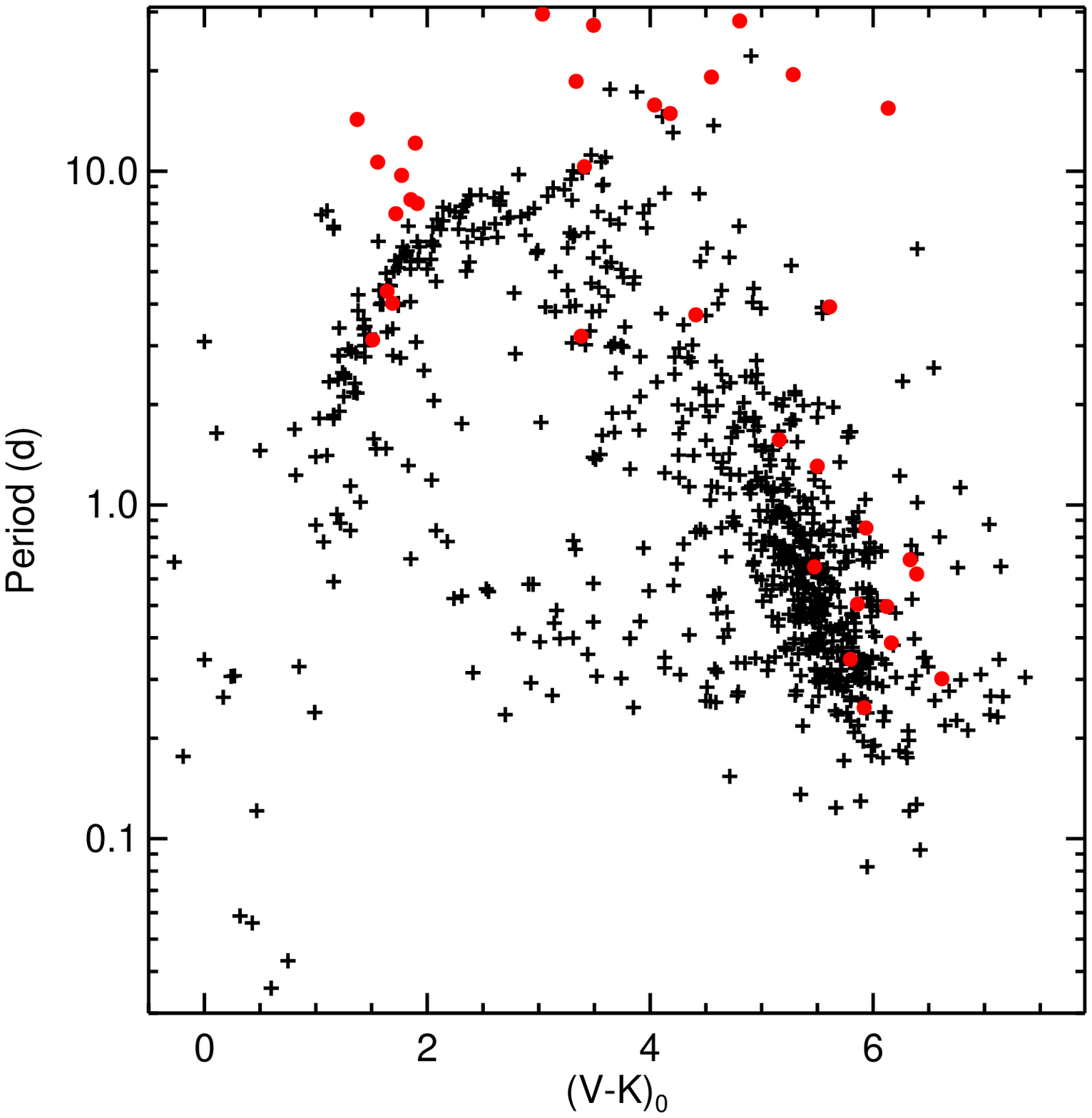}
\caption{Plot of $P$ vs.\ \vmk\ for the ensemble discussed in the rest
of the paper (black $+$) and the non-members for which we have a
period (red dots). They are more often longer periods than the rest of
the distribution at a given \vmk, consistent with being NM.}
\label{fig:pvmkNM}
\end{figure}

There are more than 150 stars where a K2 LC was obtained, presumably
because SIMBAD or other literature considered these objects as
Pleiades members. However, consideration of each of the individual
stars, including the references mentioned above in
Section~\ref{sec:membership}, suggests that these are not, in fact
likely to be Pleiades members.  They are listed in
Table~\ref{tab:nonmembers} with the period(s) we derived, and plotted
(where possible) in Fig.~\ref{fig:pvmkNM}. Only about 20\% of these
objects have periods, which is a significantly lower rate than that
for the members, consistent with these being a different population. 
Note also that several of the NM have repeated patterns that we
believe to not be spot-modulated, periodic signals; these timescales
are listed in Appendix~\ref{app:timescales}. EPIC 211145558/BPL336 (in
Fig.~\ref{fig:nmcallouts}) is discussed there.

EPIC 211094095/HII813 is one peculiar case worthy of additional
discussion. It has a very long period for a Pleiades member (19.15d;
see Fig.~\ref{fig:nmcallouts}). The LC shape is plausibly due to spot
modulation.   It has an 84\% chance of membership in Deacon \& Hambly
(2004), a 6\% chance of membership in Belikov \etal\ (1998), a 48\%
chance of membership in Stauffer \etal\ (1991), but a 98\% chance of
membership in Bouy \etal\ (2015). Its motion in RA is very discrepant
from the ensemble Pleiades motion, though the actual measurements in
the literature cover a much wider range than for other members of
comparable brightness.  We have declared it a NM based on an
unpublished HIRES spectrum (discussed in Paper III). Its radial
velocity is about 5-8 km s$^{-1}$ too high.  If it is a member, its
very long period would have to be explained in the context of models
of rotational evolution.

EPIC 211082433/V692Tau=HCG509=HHJ430 is another peculiar case. It has
two significant periods like many of the other Pleiades members; see
Fig.~\ref{fig:nmcallouts}. It appears too high in the optical CMD for
us to consider it a member. It has a 2\% chance of being a member in
Belikov \etal\ (1998), 0\% in Kazarovets (1993), 17\% in Stauffer
\etal (1991), and 16\% in Bouy \etal\ (2015). On the other hand, it is
one of two stars that  Oppenheimer \etal\ (1997) report has a high Li
abundance which means it is almost certainly young. We have opted to
leave it as a NM, but it is quite puzzling nonetheless. Interestingly,
there is a `glitch' in the phased LC at 0.374d; the other of the two
stars with Li is HCG332, discussed in Paper II as having angular dips.
It could be that this `glitch' in HCG509 is a similar effect as in
HCG332. There is no compelling evidence for IR excess in HCG509. 

\begin{deluxetable}{cccccccp{7cm}}
\tabletypesize{\scriptsize}
\tablecaption{Targets Taken as Non-Members\tablenotemark{a}\label{tab:nonmembers}}
\tablewidth{0pt}
\tablehead{\colhead{EPIC} & \colhead{RA, Dec (J2000)} & \colhead{Other name} &
\colhead{$V$ (mag)} &\colhead{\ks\ (mag)} &\colhead{\vmk} &\colhead{Period (d)}   }
\startdata
210979798 & 032928.08+223936.2 & s5035799 &   \ldots  &   11.10 &    3.03 & 
 29.564\\
210978791 & 033008.26+223838.9 & DH010 &   \ldots  &   12.77 &    4.98 & \ldots\\ 
211004869 & 033035.37+230308.3 & DH012 &   \ldots  &   14.26 &    5.33 & \ldots\\ 
210909681 & 033326.45+213229.8 & s4679029 &   \ldots  &   11.21 &    2.94
 & \ldots\\ 
210960667 & 033338.99+222108.9 & DH029 &   \ldots  &   13.44 &    4.95 & \ldots\\ 
210931896 & 033617.62+215339.1 & AKII293 &   11.00 &    9.25 &    1.64 &   4.370
\\
210953848 & 033626.15+221445.7 & DH057 &   \ldots  &   12.53 &    4.68 & \ldots\\ 
210820939 & 033629.04+201119.8 & UGCSJ033629.05+201119.6 &   \ldots  &   12.47 & 
   3.12 & \ldots\\ 
210833806 & 033753.23+202301.7 & UGCSJ033753.25+202301.2 &   \ldots  &   12.75 & 
   4.98 & \ldots\\ 
\enddata
\tablenotetext{a}{This table is available in its entirety in the online
version. A portion is shown here to demonstrate its form and content.}
\end{deluxetable}

\section{Cross-ids}

This section gives some of the common synonyms for our targets
in the literature. The table is available in its entirety in the
online version; a description of the columns appears in
Table~\ref{tab:crossids}.

\begin{deluxetable}{cl}
\tabletypesize{\scriptsize}
\floattable
\tablecaption{Contents of Online Cross-identifications List\label{tab:crossids}}
\tablewidth{0pt}
\tablehead{\colhead{Label} & \colhead{Contents}}
\startdata
EPIC & Number in the Ecliptic Plane Input Catalog (EPIC) for K2\\
Name & More common name used here \\
HII  & Name in HII catalog (Hertzsprung 1947) \\
HCG & Name in HCG catalog (Haro \etal\ 1982; Kazarovets 1993)\\
HHJ & Name in HHJ catalog (Hambly \etal\ 1993) \\
PELS & Name in Pels catalog (van Leeuwen \etal\ 1986)\\
DH & Name in DH catalog (Deacon \& Hambly 2004)\\
SRS & Name in SRS catalog (Schilbach \etal\ 1995; Belikov \etal\ 1998)\\
BPL & Name in BPL catalog (Pinfield \etal\ 2000) \\
SK & Name in SK catalog (Stauffer \etal\ 1991)\\
Tr & Name in Tr catalog (Trumpler 1921)\\
WCZ & Name in WCZ catalog (Wang \etal\ 1996)\\
Simbad & Name used in Simbad as primary identifier\\
Lodieu & Name in Lodieu \etal\ (2012) catalog\\
Bouy & Name in Bouy \etal\ (2015) catalog\\
2MASS & Name in 2MASS All-sky point source catalog\\
\enddata
\end{deluxetable}

\section{Phased LCs: Figure Set}
\label{sec:efig}

This section consists of an online-only figure set. Each periodic
member (\eg, each source in Table~\ref{tab:bigperiods}) has a PNG file
that looks like the rows of, \eg, Fig.~\ref{fig:singlefreq}. Note that
many of these sources have up to four periods that we identify; see
Paper II for more discussion of these sources. The plots in each PNG
file consist of the original light curve (in units of flux), the power
spectrum, and the phased light curve (in relative flux) for each
viable period, for up to a total of four periods.

\end{document}